\documentclass[12pt,aps,prd,nofootinbib]{revtex4-2}
\usepackage{amsmath,amsfonts,amssymb}

\usepackage{graphicx,subcaption,hyperref}
\usepackage[T1]{fontenc}
\usepackage{xcolor}
\begin{document}
\title{
Searching for heavy neutral leptons through exotic Higgs decays at the ILC
}
\author{Simon Thor}
\affiliation{KTH Royal Institute of Technology, 100 44 Stockholm, Sweden}
\author{Masaya Ishino}
\author{Junping Tian}
\affiliation{International Center for Elementary Particle Physics, The
University of Tokyo, Tokyo 113-0033, Japan}
\date{\today}

\newcommand\figref{Figure~\ref}
\newcommand\tabref{Table~\ref}
\newcommand\secref{Section~\ref}
\newcommand\equref{Equation~\eqref}

\newcommand{\ifb}[0]{\mathrm{ fb}^{-1}}
\newcommand{\pt}[0]{p_\mathrm{T}}



\begin{abstract}
In this study we investigate the feasibility of detecting heavy neutral leptons ($N_d$) through exotic Higgs decays at the proposed International Linear Collider (ILC), specifically in the channel of $e^+ e^- \to qq~ H$ with $H\to \nu N_d \to \nu~lW \to \nu l~qq$. Analyses based on full detector simulations of the ILD are performed at the center-of-mass energy of 250 GeV for two different beam polarization schemes with a total integrated luminosity of 2 $\mathrm{ab}^{-1}$. A range of heavy neutral lepton masses between the $Z$ boson and Higgs boson masses are studied. The $2\sigma$ significance reach for the joint branching ratio of $BR(H\to\nu N_d)\cdot BR(N_d\to lW)$ is about 0.1\%, nearly independent of the heavy neutral lepton masses, while the $5\sigma$ discovery is possible at a branching ratio of $0.3\%$. Interpreting these results in terms of constraints on the mixing parameters $|\varepsilon_{id}|^2$ between SM neutrinos and the heavy neutral lepton, it is expected to have a factor of 10 improvement from current constraints.
\end{abstract}
\maketitle

\section{Introduction}
The discovery of the Higgs boson at the Large Hadron Collider (LHC) \cite{ATLAS:2012yve,CMS:2012qbp} marked a monumental milestone in the field of particle physics, confirming the existence of the Higgs field and adding the final puzzle piece to the Standard Model. However, there are still gaping holes in particle physics that cannot be answered by the Standard Model, including the nature of dark matter, matter-antimatter asymmetry and more.
Physics beyond the Standard Model (BSM) is therefore a necessity. Despite this, no clear signs of BSM physics have so far been found at colliders (see Ref. \cite{lhc_searches_review} and references therein). However, the Higgs boson with its unique properties and being the least understood particle in the Standard Model, hosts great potential for being a portal to explore BSM physics.
By measuring the Higgs boson properties precisely, it is possible that BSM physics could be discovered \cite{snowmass, higgs_exotic}.

In this paper, we investigate the sensitivity of the International Linear Collider (ILC) \cite{ILC:2013jhg} in detecting heavy neutral leptons (HNL) as an exotic decay product of the Higgs boson, motivated by the model proposed in \cite{darkneutrino} to explain the matter-antimatter asymmetry problem. Our study leverages full detector simulations of the International Large Detector (ILD) \cite{ild}. We consider a range of HNL masses and branching ratios.
In the subsequent sections of this paper, we discuss the theoretical framework and models considered, outline the details of the accelerator and the detector, present the methodology employed, present the results of our analysis, and conclude with a summary of our findings and their implications.

\subsection{Theoretical framework}
There are several different theoretical models that predict heavy neutral leptons, with various physical motivations. One potential explanation for why the neutrino mass is so small compared to other particles in the Standard Model (SM) is the so-called Type-I Seesaw mechanism, which requires a heavier neutrino counterpart \cite{seesaw}. Another model is described in \cite{darkneutrino}, where baryogenesis is achieved by a model that adds a dark sector to the SM, where the first-order phase transition as well as CP-violation happen only in the dark sector and the asymmetry is converted to the SM baryon asymmetry by employing a renormalizable neutrino portal Yukawa interaction,
\begin{equation}
 \begin{aligned}
\Delta\mathcal{L}_Y = -y_{i\alpha}\bar{l}_iN_\alpha\tilde{H}+c.c.,
 \end{aligned}
\end{equation}
where $N_\alpha$ ($\alpha=u,d$) are the two singlet HNLs, $\tilde H = i\sigma_2 H^*$, where $H$ is the SM Higgs doublet, $l_i$ ($i=e,\mu,\tau$) are the SM lepton doublets, and $y_{i\alpha}$ are the corresponding Yukawa coupling constants. This Yukawa interaction generates a mixing between the SM neutrinos and HNLs with the corresponding mixing parameter $\varepsilon_{iu}$ or $\varepsilon_{id}$, which determines the coupling strength between HNLs and SM particles Higgs, $W$ and $Z$ bosons. In our study we focus on the search of $N_d$ only which is expected to have a mass at around electroweak scale, thus accessible at the ILC, while the mass of $N_u$ is expected to be much higher. Several other models also predict this interaction term between a SM neutrino, a HNL and the Higgs, with slight variations \cite{hnl_snowmass, seesaw, inverse_seesaw, neutrino_mass, neutrino_mass2}.

There are some theoretical differences between the HNL proposed in \cite{darkneutrino} compared to HNLs predicted by the seesaw mechanism. Most notably, the HNLs in the theory presented in \cite{darkneutrino} do not need to be Majorana particles, as it would not have any significant physical effect. We will therefore in this paper assume that the neutrinos are Dirac particles. From the phenomenological point of view however, there is no difference between the HNL models, as the Majorana/Dirac nature of the HNL does not have any observable effect~\footnote{The neutrinos in the dark sector in~\cite{darkneutrino} can notably decay to other dark sector particles. However this phenomenology is irrelevant for the searches in this paper.}. The results of this study are therefore applicable to all types of models that predict short-lived HNLs that decay through the decay channel that is investigated in this study.

\begin{figure}
    \centering
    \includegraphics[width=.45\linewidth]{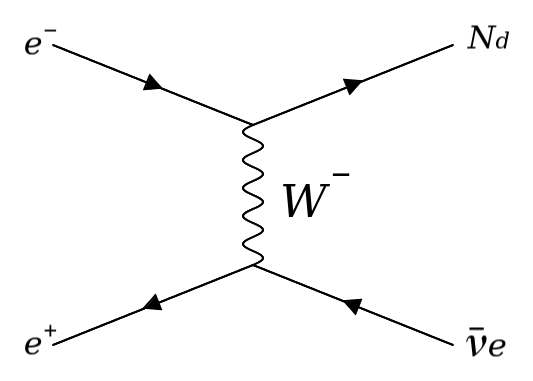}
    \caption{Direct production of HNLs through $e^+e^-$ collisions. It is also possible to have the same diagram but with swapped SM and HNLs and exchanging $\overline{\nu}_e \to \nu_e, N_d \to \overline{N}_d$.}
    \label{fig:feyn_direct}
\end{figure}

The two free parameters of the HNL that are relevant for this study are the HNL mass $m_N$ (for $N_d$) and the mixing parameter $\varepsilon_{id}$.
At colliders $N_d$ can be produced directly, as shown in \figref{fig:feyn_direct} for a representative Feynman diagram at $e^+e^-$. This would be the major method to search for the HNL when it is heavy \cite{Mekala:2022cmm}.

When its mass is below $Z$ or $W$ it is possible to search it via the decay of $Z$ or $W$. When the HNL is just heavier than $Z$ but lighter than Higgs, an interesting method is enabled by searching for the HNL from Higgs exotic decay. This is exactly the range of HNL masses that we focus on in this study. The corresponding decay widths in terms of the two free parameters are given below

\begin{align}
    \Gamma(H\to \bar\nu_j N_d) &= \left(\frac{|\varepsilon_{jd}| m_N}{v}\right)^2 \beta_f(m_H, m_N)^2 \frac{m_H}{8\pi} \equiv C_H |\varepsilon_{jd}|^2 \label{eq:h2nuNd}\\
    \Gamma(N_d \to l_i^- W^+) &= \frac{(|\varepsilon_{id}| g)^2}{64\pi}\beta_f(m_N, m_W)^2 \frac{m_N^3}{m_W^2} \left(1 + 2\left(\frac{m_W}{m_N}\right)^2\right) \equiv C_W |\varepsilon_{id}|^2 \label{eq:Nd2lW} \\
    \Gamma(N_d\to \nu_i Z) &= \frac{(|\varepsilon_{id}| g)^2}{128\pi} \beta_f(m_N, m_Z)^2 \frac{m_N^3}{m_W^2} \left(1 + 2\left(\frac{m_Z}{m_N}\right)^2\right) \equiv C_Z|\varepsilon_{id}|^2 \label{eq:Nd2nuZ}
\end{align}

Here, $v$ is the vacuum expectation value 246 GeV, $\beta_f(M, m) = 1-(m/M)^2$, $g$ is the electroweak $SU(2)$ coupling constant, $m_W$ is the W boson mass, and $m_Z$ is the Z boson mass, and $i, j$ indicate the lepton flavor. The equations are slightly modified versions of the ones given in \cite{nd_higgs}. We have defined $C_H, C_W, C_Z$ as the product of all terms that depends on the HNL mass instead of the mixing parameter. The charge conjugate decay mode $H\to \nu_j \bar N_d$ is of course also possible, with the corresponding decays $\bar N_d \to l^+_i W^-$ and $\bar N_d \to \bar \nu_i Z$ and same decay widths. The decay $N_d \to l^+_i W^-$ is however not possible if the HNL is a Dirac particle (which we assume).
In other literature (such as in Ref. \cite{nd_higgs}), the mixing matrix is sometimes referred to as $V$ or $U$.

\begin{figure}
    \centering
    \begin{subfigure}[t]{.52\linewidth}
        \centering
        \includegraphics[width=\linewidth]{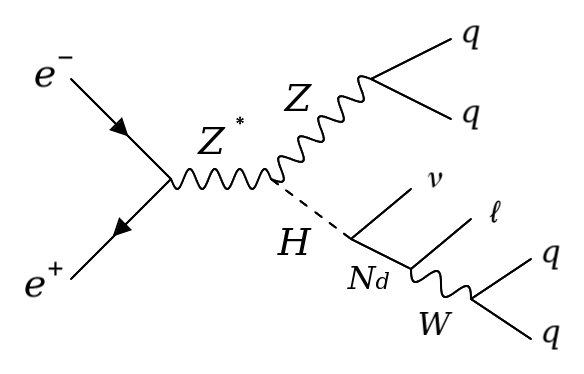}
    \end{subfigure}
    \hfill
    \begin{subfigure}[t]{.47\linewidth}
        \centering
        \includegraphics[width=\linewidth]{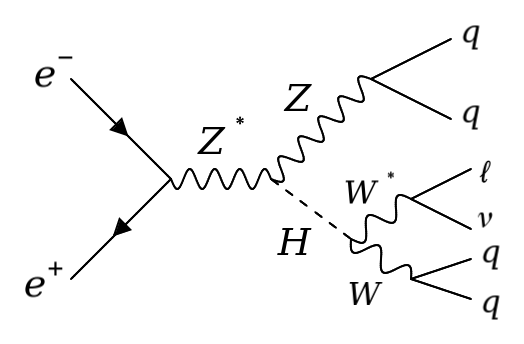}
    \end{subfigure}
    \caption{Feynman diagrams for the signal (left) and the main background (right) in this study.}
    \label{fig:sig_bkg_feyn}
\end{figure}

Given above theory basis we are ready to define our signal process at the ILC. We will take advantage of the leading Higgs production channel (Higgs-strahlung process) $e^+e^-\to ZH$ and look for the Higgs exotic decay mode $H\to\bar{\nu}N_d$. We will concentrate on the dominant decay channel where $Z\to q\bar{q}$ and $N_d\to l^-W^+\to l~q\bar{q}$. The charge conjugate channel is also targeted as our signal process. The Feynman diagram of this signal process is shown in \figref{fig:sig_bkg_feyn} (left). The observable will be the event rate of the signal process, which is basically the product of the cross section of $e^+e^-\to ZH$ cross section ($\sigma_{ZH}$) and decay branching ratios ($BR$) of $Z\to q\bar{q}$, $H \to \nu N_d$, $N_d \to l^-W^+$ and $W\to q\bar{q}$. With $\sigma_{ZH}$, $BR(Z\to q\bar{q})$ and $BR(W\to q\bar{q})$ precisely measured in other processes at the ILC, the observable here essentially becomes a joint branching ratio of $H$ and $N_d$ decays, $BR(H\to \bar\nu N_d)\cdot BR(N_d\to l^- W^+)$, which can be computed as a function of the two free parameters $m_N$ and $\varepsilon_{id}$ as follows (using equations \ref{eq:h2nuNd}, \ref{eq:Nd2lW}, \ref{eq:Nd2nuZ})

\begin{align}
    &\sum_j BR(H\to \bar\nu_j N_d)\cdot BR(N_d\to l_i^- W^+) \nonumber \\
    &= \frac{\sum_j\Gamma(H\to \bar\nu_j N_d)}{\Gamma_{SM} + \sum_j\Gamma(H\to \bar\nu_j N_d)} \frac{\Gamma(N_d \to l_i^- W^+)}{\sum_k (\Gamma(N_d\to l_k^- W^+) + \Gamma(N_d \to \nu_k Z))} \nonumber \\
    &= \frac{\sum_j |\varepsilon_{jd}|^2 C_H}{\Gamma_{SM} + \sum_j |\varepsilon_{jd}|^2 C_H} \frac{C_W |\varepsilon_{id}|^2}{\sum_j (|\varepsilon_{jd}|^2 C_W + |\varepsilon_{jd}|^2 C_Z)} = \frac{C_H}{\Gamma_{SM} + \sum_j |\varepsilon_{jd}|^2 C_H} \frac{C_W |\varepsilon_{id}|^2}{C_W + C_Z} \nonumber \\
    &\approx \frac{C_H}{\Gamma_{SM}} \frac{C_W }{C_W + C_Z}|\varepsilon_{id}|^2, \label{eq:br2eps}
\end{align}
where the approximation in the last step holds when the decay width contribution to the Higgs from the HNL is far smaller than the Higgs SM decay width $\Gamma_{SM}$. Equation~\ref{eq:br2eps} tells that the observable defined above for the channel with $l_i$ is proportional to the corresponding mixing parameter squared $\varepsilon_{id}$. More explicitly, when the charged lepton in the final state is $e$ the observable depends simply on $\varepsilon_{ed}$ instead of a combination of $\varepsilon_{ed}$, $\varepsilon_{\mu d}$ and $\varepsilon_{\tau d}$. Numerically, for our interested mass range $m_Z<m_N<m_H$, $\frac{C_W}{C_W+C_Z}$ is $O(1)$ ($>75\%$, closer to 1 for smaller $m_N$), and $\frac{C_H}{\Gamma_{SM}}$ is $O(10)$.

\subsubsection{Current constraints} \label{sec:current_constraints}
It's worth summarizing here current constraints on the two free parameters in above model. First of all as given in \cite{darkneutrino}, based on the test of lepton universality \cite{Pich:2013lsa}$ |\varepsilon_{id}|^2$ is typically constrained to $|\varepsilon_{id}|^2\lesssim 10^{-3}$. This constraint is independent of the HNL mass. In the mass range below $m_Z$, the constraint from $Z$ decay comes from DELPHI at LEP \cite{delphi}. Newer searches by the ATLAS and CMS collaborations have instead investigated decays of $W\to l N_d$, which put similar or slightly stronger constraints on the mixing parameters as the DELPHI search \cite{atlas_trilepton, cms_trilepton}.

For short-lived heavy neutral leptons, when $m(N_d)$ is between 5 and 50 GeV, the constraint can be as strong as $\varepsilon\lesssim 2\times 10^{-5}$, while it is weaker when $m_N>50$ GeV. The constraints from searches of heavy neutral leptons at the LHC experiments \cite{CMS:2018iaf, ATLAS:2019kpx} can also be cast into the parameters defined here, notably in the mass range well below $m_W$ ($\lesssim 10$ GeV) the limit on $|\varepsilon_{id}|^2$ can be as strong as $10^{-6}$ if the heavy neutrino is long-lived and therefore has a displaced vertex. 

For the mass range we are interested in this study, there is no stronger limit yet from current LHC experiments. According to \cite{nd_higgs}, the measurement of Higgs total width at the LHC can set an indirect limit which is very weak at this moment; stronger limit might be possible using Higgs exotic decay with a leptonic final state $H\to \nu N_d \to ll\nu\nu$, based on DELPHES fast detector simulation. 
However, the constraints from this theory assumes that the HNL only mixes with one SM neutrino, which for a general theory is not true. This, in combination with the fact that fast detector simulations were used, means that the constraints from Ref. \cite{nd_higgs} are likely optimistic.

It is interesting to note another method which can give indirect limit. At parton level, Higgs SM decay to $WW^*$ can give exactly same final states as Higgs exotic decay to $\nu N_d$, illustrated in \figref{fig:sig_bkg_feyn} (right). Thus the measurement of $H\to WW^*$ branching ratio can provide a constraint on $BR(H\to \nu N_d)$ if the event selection efficiency for $H\to\nu N_d$ events is not zero in the $H\to WW^*$ analysis. The current measurements of the branching ratio is $BR(H\to WW^*) = 25.7 \pm 2.5\%$ \cite{atlas_higgs_br}. Assuming that all of $H\to \nu N_d$ decays contribute to the $H\to WW^*$ decay channel, the $2\sigma$ limit for the branching ratio of $H\to \nu N_d$ is $5\%$. This constraint on the branching ratio is overly optimistic but is nevertheless included as a comparison.

All of the above mentioned constraints are shown in the results section and compared to the ILC constraints.

\section{Simulation framework}

\subsection{International Linear Collider}
The International Linear Collider (ILC) is a proposed future linear $e^+e^-$ collider. One of the main goals of the collider is to be a Higgs factory, i.e., produce Higgs bosons and perform precision measurements of its properties. The hope is that some of these properties will deviate from predictions by the SM and will therefore be a hint of BSM, as explained earlier. The center-of-mass energy is planned to be 250 GeV at the start, with possibilities to extend the accelerator and thus increasing the collision energy in later stages.
At 250 GeV, the cross section for Higgsstrahlung process, i.e., $e^+e^- \to ZH$ (see Feynman diagrams in \figref{fig:sig_bkg_feyn}) reaches its maximum value and is therefore a suitable center-of-mass energy to perform measurements of the Higgs. The latest overviews of ILC can be found in the ILC documents for the Snowmass 2022 \cite{snowmass} and European Strategy Update for HEP 2020 \cite{Bambade:2019fyw}. Our study is assuming a total integrated luminosity of 2 $\mathrm{ab}^{-1}$ at the ILC $\sqrt{s}=250$ GeV, equally shared among two beam polarization schemes: $P(e^-,e^+)=(-0.8,+0.3)$ and $P(e^-,e^+)=(+0.8, -0.3)$. 
While the exact beam polarization configurations have not been decided, the suggested beam polarization in the European Strategy Update \cite{Bambade:2019fyw} closely resembles this (with some minor contributions of $(+0.8, +0.3)$ and $(-0.8, -0.3)$ which we ignore for simplicity and since they will have such a small signal).

\subsection{The ILD concept}
Our full detector simulation is based on the ILD which is one of the two proposed detectors for the ILC. It has a hybrid tracking system and highly granular calorimeters optimized for Particle Flow reconstruction. It has been developed to be optimized for precision measurements of the Higgs boson, the weak gauge bosons and the top-quark, as well as for direct searches of new particles \cite{ild}. The subdetectors relevant for this study are briefly described below.

The vertexing system consists of three double layers of silicon pixel detectors and is located closest to the interaction point. 
The hybrid tracking system consists of both a silicon strip detector and a time projection chamber (TPC). 
The whole tracking system is located outside the vertexing system. 
The electromagnetic calorimeter (ECAL), located outside the tracking system, is a sampling calorimeter made out of silicon and tungsten in finely segmented pads of $(5\times5) \mathrm{~mm}^2$. 
The hadronic calorimeter is located outside the ECAL and is based on scintillator as default consisting of $(3\times3) \mathrm{~cm}^2$ tiles.
Outside of the calorimeters there is a superconducting solenoid with a magnetic field of $3.5 \mathrm{~T}$. An iron return yoke located outside the coil works as a muon identification system. 

\subsection{Software}
This study utilizes the software package ILCSoft v02-02 \cite{ilcsoftDoc} to conduct simulations and reconstructions. The parameters of the incoming beams are simulated using the GUINEA-PIG package \cite{beamstrahlung, guinea-pig}. The beam spectrum, including beamstrahlung and initial state radiation (ISR), is taken into account.
In line with the current ILC design, the beam crossing angle of 14 mrad is taken into consideration.
For the generation of Monte Carlo (MC) samples of the signal and SM background events, the WHIZARD 2.8.5 event generator \cite{whizard, omega} is employed. The signal events were generated by employing the UFO model that was developed in \cite{Mekala:2022cmm}, with 6 different values for HNL mass $m_N=95,100,105,110,115,120$ GeV. The parton shower and hadronization model is adopted from PYTHIA 6.4 \cite{pythia6.4}.
To simulate the detector response, the generated events are passed through the ILD simulation \cite{ild} (model version ILD\_l5\_v02) implemented with the DD4HEP \cite{Frank_2014,Frank_2015} software package, which is based on Geant4 \cite{geant4, geant4dev, geant4app}. Event reconstruction is performed using the Marlin \cite{marlin} framework. The PandoraPFA \cite{pfa} algorithm is specifically employed for calorimeter clustering and the analysis of track and calorimeter information, following the particle flow approach.

The samples used for the background are all SM processes (excluding ones with the Higgs boson) where two, four, and six fermions are produced. Additionally, processes where two quarks and one Higgs boson are produced (almost exclusively $e^+ e^- \to ZH \to q\bar{q} H$) are included as background, including all possible SM decay modes of the Higgs. Each background can also be separated into leptonic (only leptons in the final state), semileptonic (leptons and hadrons in the final state) or hadronic (only hadrons in the final state) decay channels.

The three-fermion final state where the incoming electron or positron interacts with a photon was also investigated as a background. However, all simulated electron-gamma samples were excluded after the cuts applied (see further down) and are therefore not included in the background.
The cross sections for the background processes are given in \tabref{tab:bkg_xsecs}.

\begin{table}[h]
\setlength\tabcolsep{2pt}
    \centering
    \begin{tabular}{l|l|r|r}
        Process & Abbreviation & Cross section $(-0.8, +0.3)$ [fb] & Cross section $(+0.8, -0.3)$ [fb] \\
        \hline
        2 fermion leptonic & 2f\_l & 13 000 & 10 300 \\
        2 fermion hadronic & 2f\_h & 77 300 & 45 700 \\
        4 fermion leptonic & 4f\_l & 10 400 & 6 110 \\
        4 fermion semileptonic & 4f\_sl & 19 200 & 2 840 \\
        4 fermion hadronic & 4f\_h & 16 800 & 1 570 \\
        6 fermion & 6f & 1.28 & 0.26 \\
        $e^+ e^- \to q\bar{q} H$ & qqh & 208 & 140 \\
        Signal (BR = 1\%) & Signal & 1.396 & 0.941
    \end{tabular}
    \caption{Cross sections of the various background processes \cite{elog}. For example, ``2 fermion hadronic'' means that there are two hadronic fermions produced in the final state, i.e., all SM processes (excluding processes involving the Higgs) of $e^+e^- \to q\bar q$. The signal cross section shows the case when the branching ratio is 1\%.}
    \label{tab:bkg_xsecs}
\end{table}

The analysis for applying cuts were done with the ROOT C++/Python framework 6.28 \cite{root} and Jupyter notebooks \cite{jupyter}. The machine learning model was developed with TMVA \cite{tmva}.

\section{Analysis}

For each polarization scheme and each mass value of HNL, the event selection and cuts applied to reduce the background were done in three stages: pre-selection, rectangular cuts, and a machine learning cut, explained in detail in the following subsections. As a preview of the analysis procedure a cut flow table of all the cuts applied and the number of events that pass each cut, separated for the different background categories, are shown in \tabref{tab:cuttable}. The table is an example of the cuts for a HNL mass of 100 GeV, a joint branching ratio of 1\%, beam polarization of $(+0.8, -0.3)$, and an integrated luminosity of 1 $\mathrm{ab}^{-1}$.

\begin{table}[h]
\setlength\tabcolsep{2pt}
    \centering
    \begin{tabular}{l|r|r|r|r|r|r|r|r|r|r}
        Cut & Signal & Background & $\sigma$ & 2f\_l & 2f\_h & 4f\_l & 4f\_sl & 4f\_h & 6f & qqh \\
        \hline
        No cuts & 941 & 66651497 & 0.12 & 10314870 & 45672588 & 6114301 & 2839022 & 1570051 & 260 & 140405 \\
        Pre-selection & 831 & 12565351 & 0.23 & 5696748 & 979693 & 4109167 & 1739683 & 22431 & 194 & 17434 \\
        cut 1 & 769 & 1287215 & 0.68 & 70332 & 146740 & 897907 & 149918 & 15416 & 142 & 6759 \\
        cut 2 & 722 & 1025729 & 0.71 & 61382 & 49161 & 785129 & 120467 & 4506 & 132 & 4952 \\
        cut 3 & 708 & 434591 & 1.07 & 44787 & 22077 & 293992 & 67433 & 2031 & 74 & 4197 \\
        cut 4 & 665 & 24666 & 4.18 & 399 & 4093 & 1176 & 13462 & 1687 & 72 & 3777 \\
        cut 5 & 583 & 6919 & 6.73 & 0 & 1151 & 0 & 1234 & 1384 & 55 & 3094 \\
        cut 6 & 574 & 4487 & 8.07 & 0 & 544 & 0 & 666 & 648 & 19 & 2611 \\
        MVA cut & 434 & 1162 & 10.87 & 0 & 52 & 0 & 26 & 79 & 6 & 999
    \end{tabular}
    \caption{Cut flow table for a HNL mass of 100 GeV, a joint branching ratio of 1\%, beam polarization of $(+0.8, -0.3)$ and an integrated luminosity of 1 $\mathrm{ab}^{-1}$. The numbered cuts represent rectangular cuts, which are explained in more detail below. The column named $\sigma$ is the signal significance in units of standard deviation $\sigma$.}
    \label{tab:cuttable}
\end{table}

\subsection{Pre-selection}
The final state of signal events consists of one charged lepton, one neutrino and four jets. The pre-selection is applied to reconstruct the basic information of the leptons and jets, as well as to properly pair them into $W$, $Z$, $N_d$ and Higgs in each event, based on the signal characteristics. The pre-selection will supply the necessary information for the next cuts. The procedure of pre-selection is briefly explained here. 

Isolated leptons were first identified in each event using a pre-trained neural network by the IsolatedLeptonTagging algorithm in iLCSoft \cite{marlin}. 
As a brief explanation of this algorithm, the isolated leptons are required to have a momentum of at least 5 GeV; the energies deposited in ECAL and HCAL in comparison with their momentum are required to match the pattern of electron or muon; the impact parameters of the leptons are required to be consistent with particles from primary interaction point; a double cone is defined around each candidate lepton and the amount of particles in each layer of the cone are taken into account to distinguish the isolated leptons and the leptons from jets; more details about exact input variables and their distributions can be found in~\cite{Tian:ISOLep}. 
The neural network gives a numerical output for each particle of the event usually between 0 an 1 (though it can give a higher value, even up to 2), with a higher value meaning that a particle is more likely an isolated lepton. If the particle is a muon, it is required that the isolated lepton finder output is greater than 0.5, whereas if it is an electron, it is required to be greater than 0.2. Only a loose cut is applied on this parameter in the pre-selection not to remove signal and background events too early.
Each event is then required to have at least one isolated lepton according to these criteria. For the signal, typically only one lepton fulfills this but there are occasionally ($<1\%$) 2 or more leptons. In that case, the highest energy lepton is chosen as the isolated lepton.

\begin{figure}
    \centering
    \includegraphics[width=.6\linewidth]{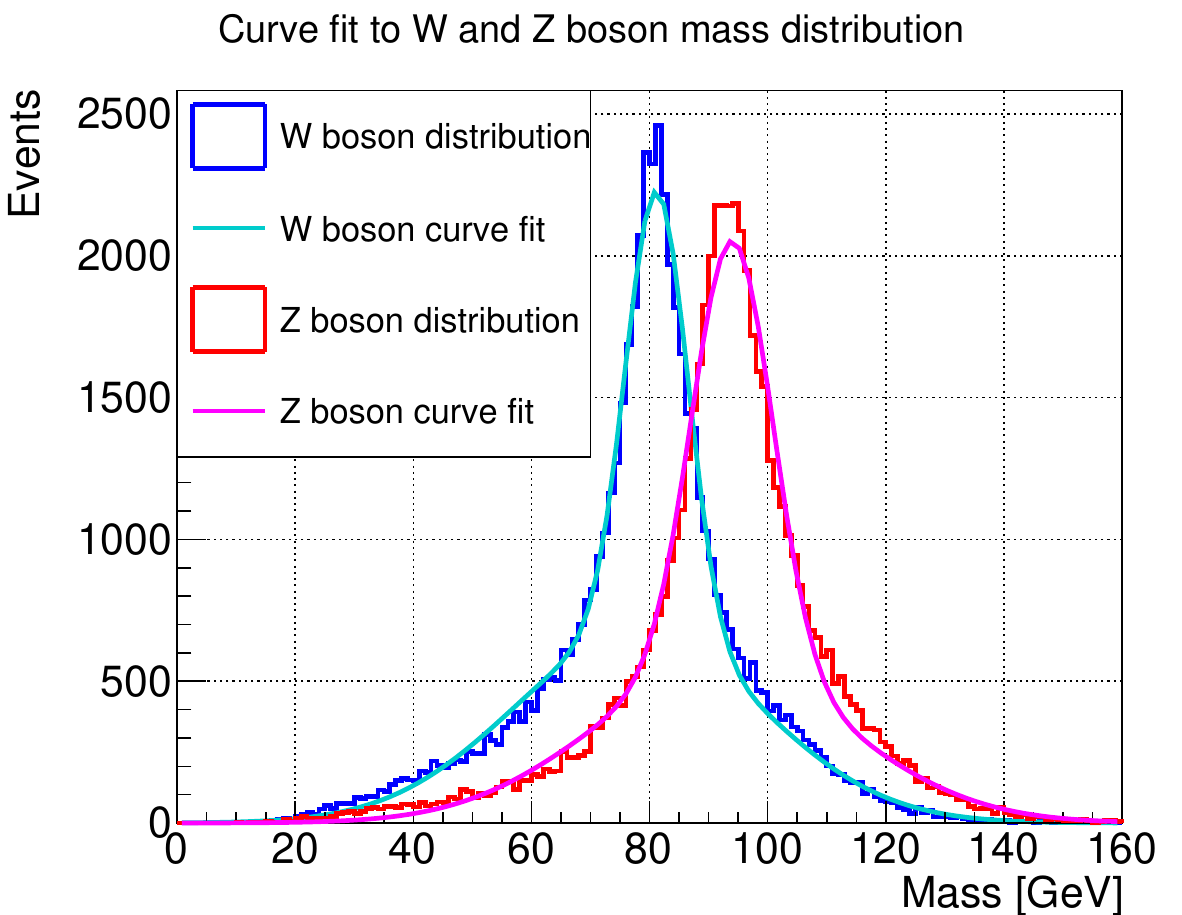}
    \caption{Reconstructed mass distributions of the W and Z bosons, by pairing jets using MC truth information. The smooth lines are curve fits to the histograms.}
    \label{fig:massres}
\end{figure}

The remaining particles in the event are clustered into four jets using the Durham jet clustering algorithm \cite{Durham}. If this fails, the event is also rejected. The four jets are then paired with each other and classified as $W$ jets or $Z$ jets based on which pairing minimized
\begin{equation}
    \chi^2 = \left(\frac{m_W - m_{12,jet}}{\Delta m_{W,jet}}\right)^2 + \left(\frac{m_Z - m_{34,jet}}{\Delta m_{Z,jet}}\right)^2 .
\end{equation} 
Here, $m_W = 80.4$ GeV and $m_Z = 91.187$ GeV are the W and Z boson masses. $m_{12,jet}, ~m_{34,jet}$ are the reconstructed masses based on a certain pairing of jets and their 4-momenta. $\Delta m_{W,jet} = 5.3~\mathrm{GeV}$, $\Delta m_{Z,jet} = 6.8$ GeV represent the mass resolutions of $W$ and $Z$ bosons. 
This mass resolution was calculated by using MC truth information to identify which of the reconstructed jets originate from $W$ or $Z$ jets
\footnote{To identify which jet corresponded to which boson, the reconstructed constituent particles in each jet were linked to their corresponding MC truth particle, and by traversing the tree of interactions it could be determined if a certain jet constituent particle originated from a $W$ or $Z$ boson. The scalar energy sum of all constituent particles originating from a $Z$ and $W$ boson respectively was calculated. If most of the energy came from the $Z$ boson, the jet was classified as a $Z$ jet and likewise for $W$ jets. If more than two jets were classified as either $Z$ or $W$ jets, the event was ignored.}.
After each jet was classified as a $W$ or $Z$ jet, the jet pair 4-momenta were added and the reconstructed masses were computed. The mass resolution for each boson was then set as the standard deviation of the mass distribution, when performing a fit of a normal distribution to the mass distributions. 
The mass distributions used are shown in \figref{fig:massres}, for $e^+e^- \to ZH \to q\bar{q}~WW^* \to q\bar{q}~q\bar{q}~l\nu$ events. Since this process is the dominant background, it has the same final state and has very similar kinematics as the HNL model, the $W$ and $Z$ mass resolutions are similar to signal events.

After the jet clustering and jet pairing, the $W$ and charged lepton are grouped to form reconstructed $N_d$. The missing 4-momentum, calculated as the 4-momentum of initial state minus the total 4-momentum of all reconstructed visible particles in the final state, is reconstructed as the 4-momentum of $\nu$. Then Higgs is reconstructed as sum of the 4-momenta of $N_d$ and $\nu$. After the pre-selection most of the hadronic background events are rejected, as they do not have an isolated lepton, and large portions of the leptonic backgrounds are also reduced, as they fail the 4-jet reconstruction; details shown in \tabref{tab:cuttable}. 

\subsection{Rectangular cuts}
For the rectangular cuts, various observables were identified by comparing their distributions between signal and background events. Cut values were optimized to maximize the final signal significance after the rectangular cuts and the machine learning cut.

\begin{figure}
    \centering
    \begin{subfigure}[t]{.49\linewidth}
        \centering
        \includegraphics[width=\linewidth]{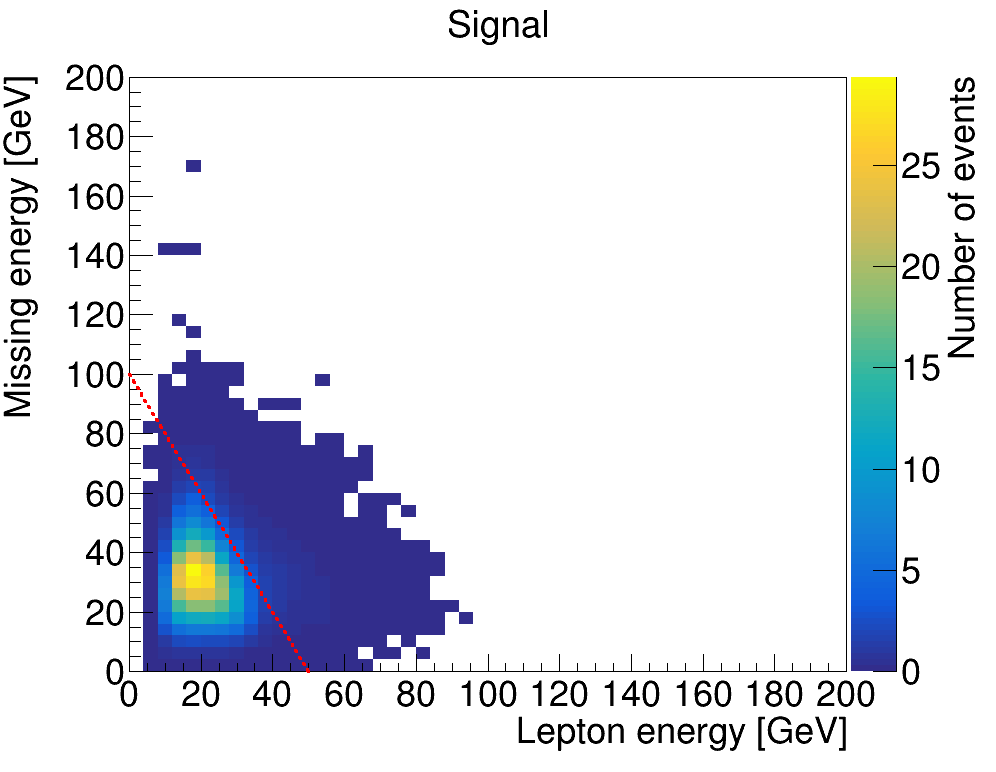}
    \end{subfigure}
    \hfill
    \begin{subfigure}[t]{.49\linewidth}
        \centering
        \includegraphics[width=\linewidth]{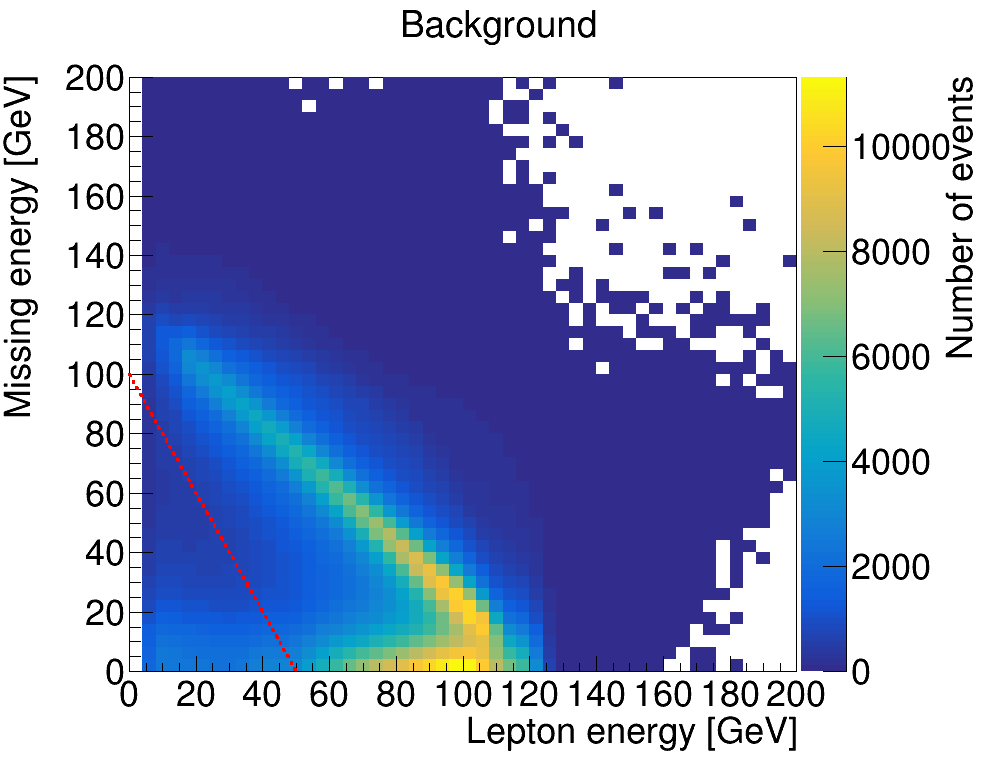}
    \end{subfigure}
    \caption{2D distributions of the signal and background for isolated lepton energy ($x$-axis) and missing energy ($y$-axis) for a beam polarization of $(+0.8, -0.3)$ and a HNL mass of 100 GeV. The background distribution shown is the 4f\_sl background, where the red line shows the cut that is applied. The number of events shown for the signal is for a branching ratio of 1\% but the distribution does not change for other branching ratios.}
    \label{fig:elep_emis}
\end{figure}

The first cut applied was a combination of missing energy and lepton energy. This was mainly to reduce the leptonic and semileptonic backgrounds, which have high energy leptons or neutrinos.
As an example, the 2D distributions of the lepton energy ($E_{lep}$) and missing energy ($E_{mis}$) for the 4f\_sl background and signal are shown in \figref{fig:elep_emis}. The plot (and all subsequent plots in this section) are for a beam polarization of $(+0.8, -0.3)$, HNL mass of 100 GeV, and a branching ratio of 1\% for the signal. The cut (cut 1) was set to $E_{lep}/50\mathrm{GeV} + E_{mis}/100\mathrm{GeV} < 1$. The results of al cuts shown in figures below, are the ones shown in \tabref{tab:cuttable}.

\begin{figure}
    \centering
    \includegraphics[width=.6\linewidth]{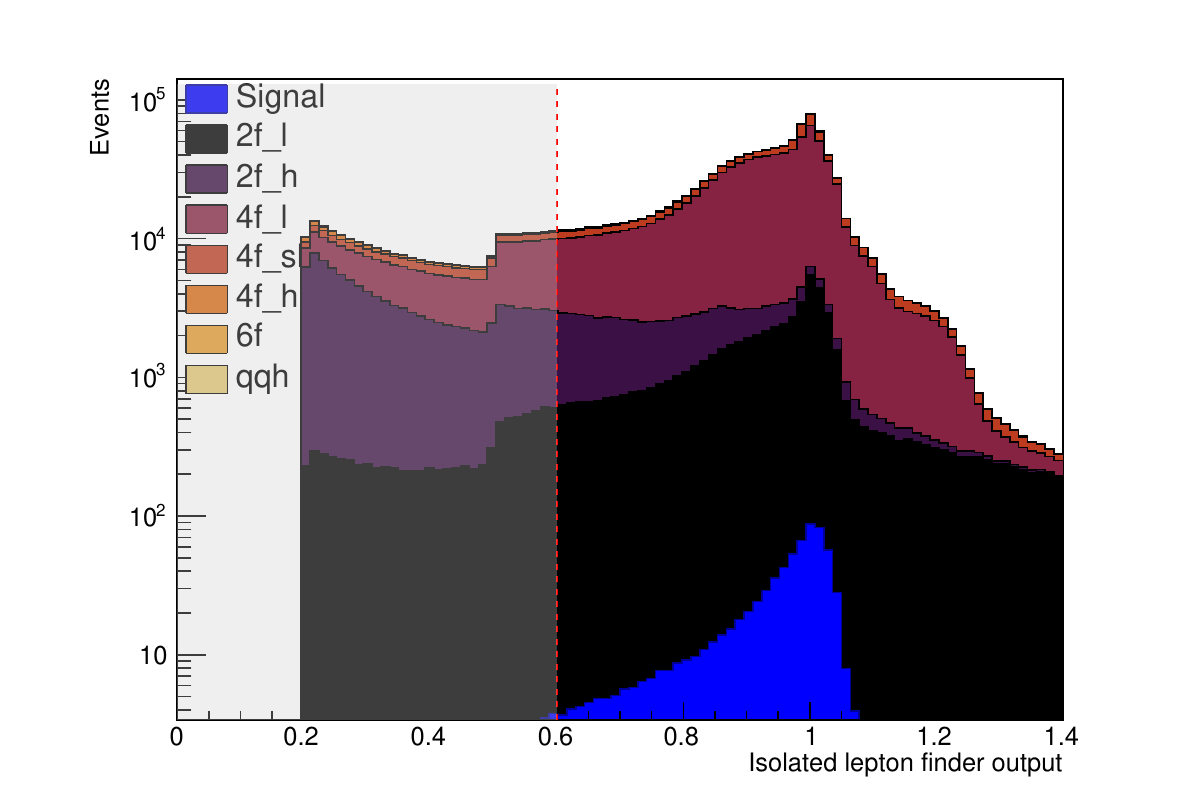}
    \caption{The distributions of the isolated lepton finder output for the signal and background, where different colors indicate different background processes. The same beam polarization and HNL mass is shown as in \figref{fig:elep_emis}. The distributions are properly normalized according to the corresponding cross sections. The red vertical line shows the cut value and the grey region is the region that is rejected. A logarithmic scale is used on the $y$-axis.}
    \label{fig:mvalep}
\end{figure}

The second cut applied was on the isolated lepton finder output, which is required to be greater than 0.6 (cut 2). This cut is tighter than the loose cut applied by default in the isolated lepton finder algorithm in the pre-selection.
This cut mainly suppresses further the hadronic backgrounds (2f\_h and 4f\_h) where a lepton from a jet was mistagged as an isolated lepton. The distributions for the signal and background, as well as the cut value is shown in \figref{fig:mvalep}. 
Note that the events shown are only the events passing the previous cut(s). This applies to all the upcoming similar plots in this section.

\begin{figure}
    \centering
    \includegraphics[width=.6\linewidth]{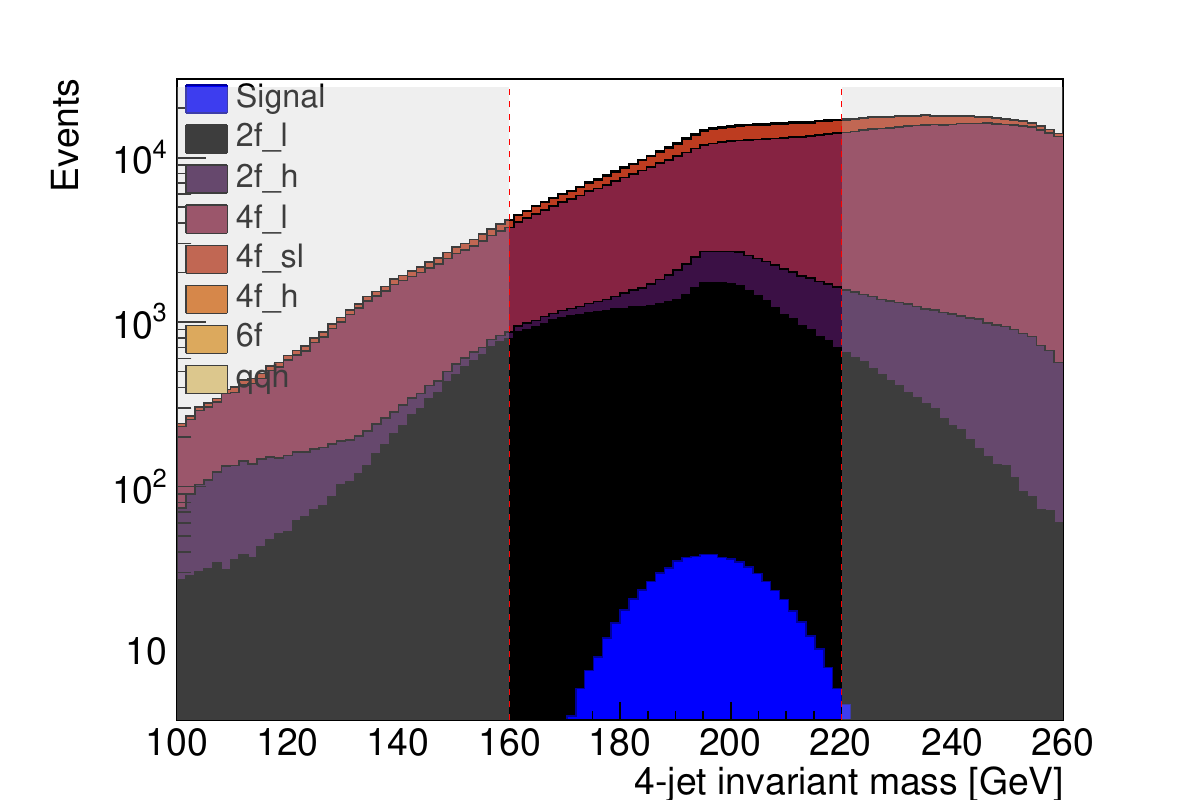}
    \caption{The distributions of 4-jet invariant mass for the signal and background. The figure format is the same as in \figref{fig:mvalep}.}
    \label{fig:mjet}
\end{figure}

Next, the 4-jet combined invariant mass is required to be between 160 and 220 GeV (cut 3), which reduces both the hadronic (at the high-mass region) and semi-leptonic / leptonic (at the low-mass region) backgrounds. 
The distributions of the 4-jet invariant mass is shown in \figref{fig:mjet} for the signal and the total background events.

\begin{figure}
    \centering
    \includegraphics[width=.6\linewidth]{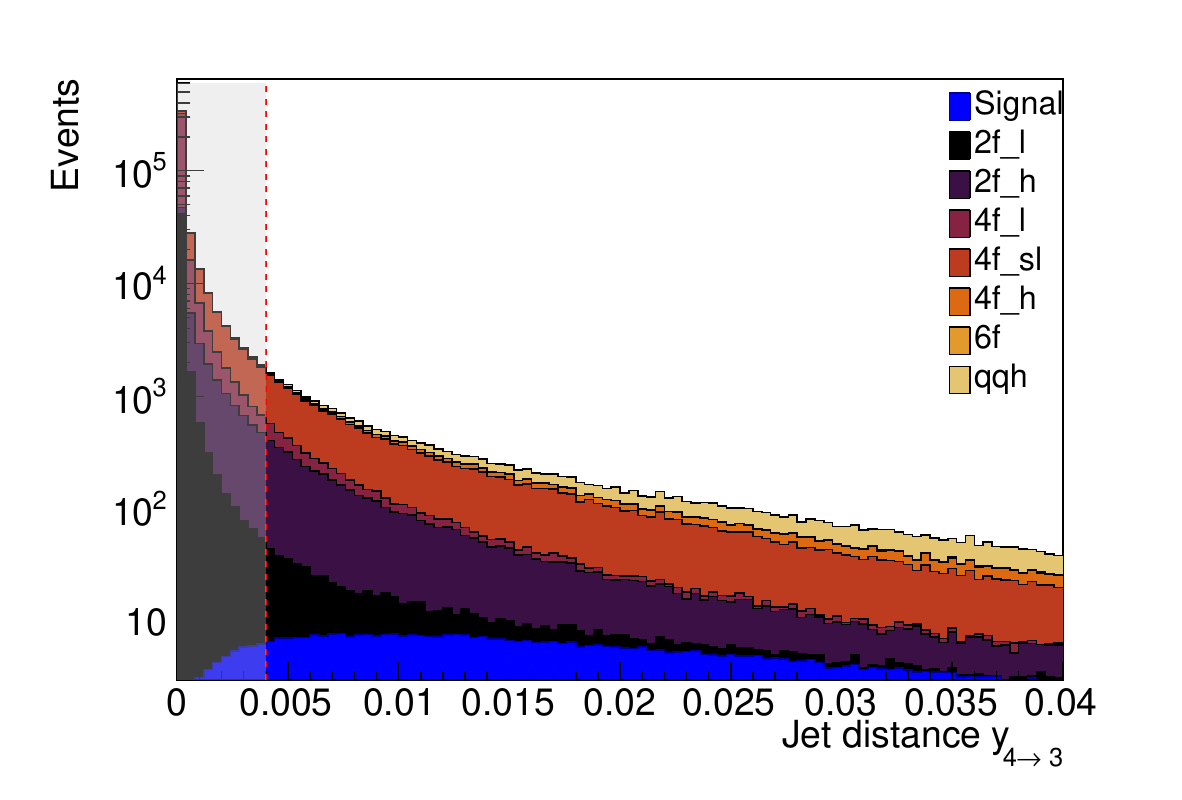}
    \caption{The distributions of the jet distance $y_{4\to3}$ for the signal and background. The figure format is the same as in \figref{fig:mvalep}.}
    \label{fig:y34}
\end{figure}

The fourth cut applied is on the Durham jet distance $y_{4\to3}$, calculated as
\begin{equation}
    y_{4\to3} = \min_{i,j} \left\{ \frac{2\min\{E_i, E_j\}^2 (1-\cos\theta_{ij})}{E_{vis}^2} \right\} ,
\end{equation}
where $E_i$ is the energy of jet $i$, $\theta_{ij}$ is the angle between jets $i$ and $j$, and $E_{vis}$ is the total energy of the four jets. This jet distance is used for the jet clustering and the pair of jets that give the smallest jet distance are combined, hence the use of $\min$ in the equation above. This clustering is performed multiple times until there are only four jets left, and $y_{4\to3}$ is the minimum jet distance at this stage. 
By requiring $y_{4\to3} > 0.004$ (cut 4), this can therefore help filter out semi-leptonic and leptonic background events, as shown in \figref{fig:y34}.

\begin{figure}
    \centering
    \includegraphics[width=.6\linewidth]{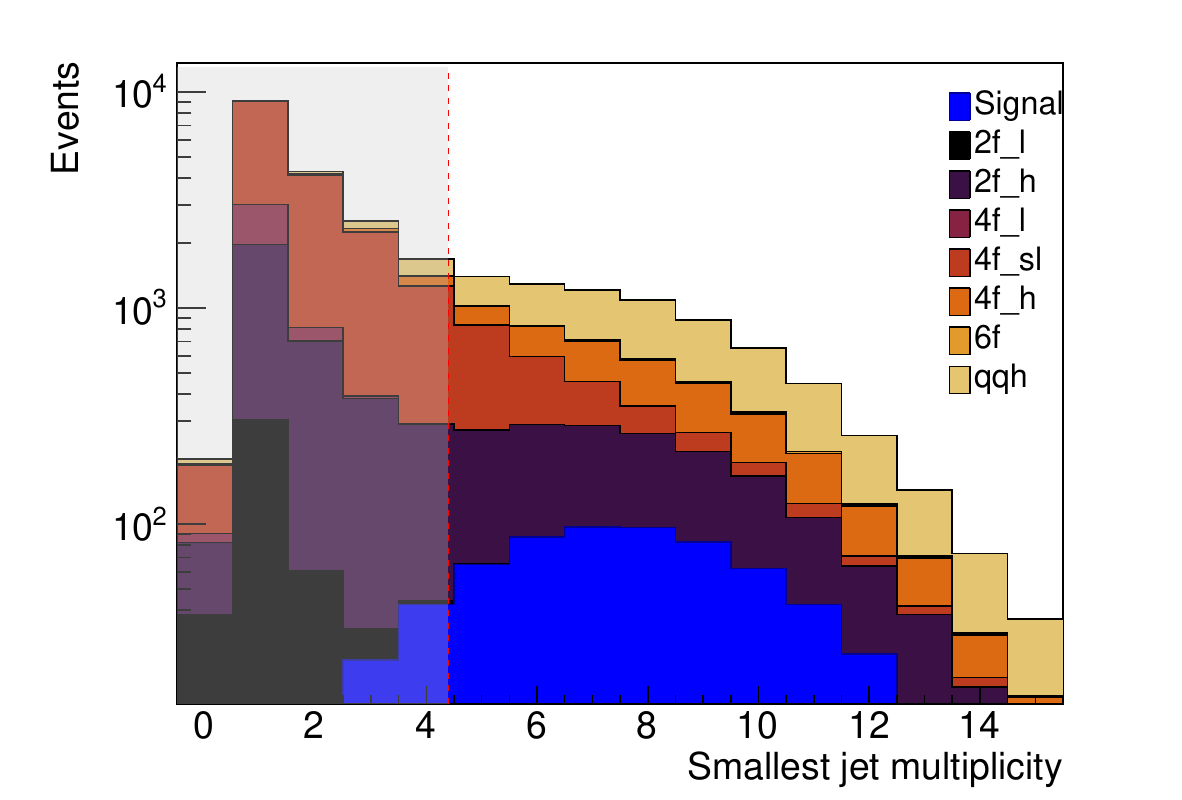}
    \caption{The distributions of the smallest number of particles in the jets of each event. for the signal and background. The figure format is the same as in \figref{fig:mvalep}. The histogram bins are centered at each integer.}
    \label{fig:nparticles}
\end{figure}

The fifth cut applied is that any jet in the event must have at least four particles (cut 5). 
This cut further reduces the semi-leptonic and leptonic backgrounds, as shown in \figref{fig:nparticles}.

\begin{figure}
    \centering
    \begin{subfigure}[t]{.49\linewidth}
        \centering
        \includegraphics[width=\linewidth]{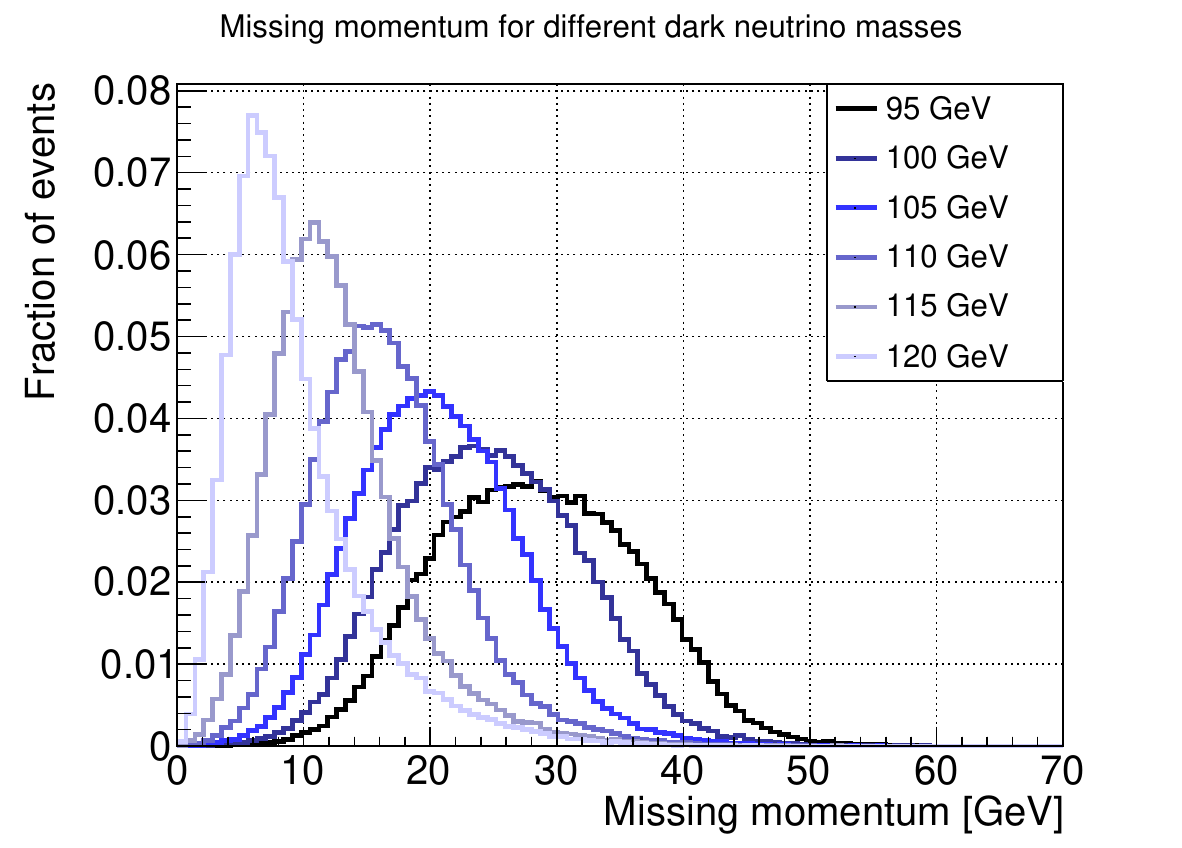}
    \end{subfigure}
    \hfill
    \begin{subfigure}[t]{.49\linewidth}
        \centering
        \includegraphics[width=\linewidth]{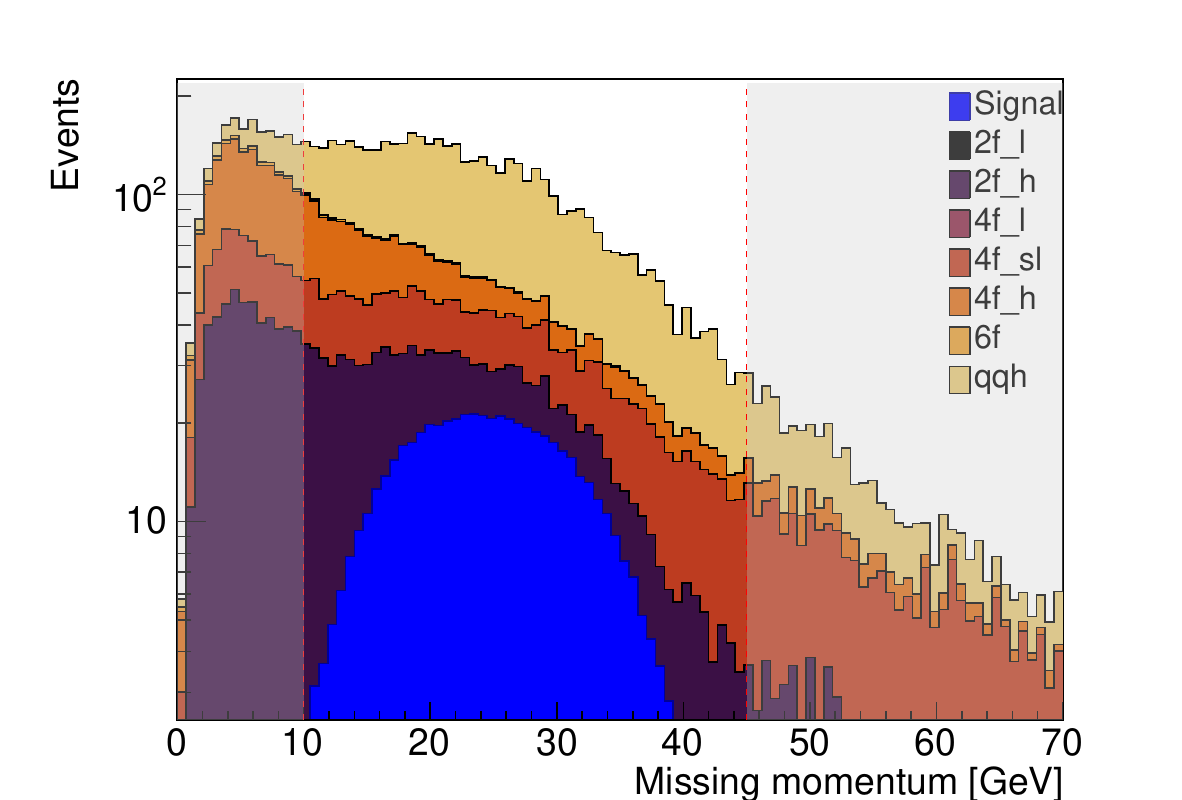}
    \end{subfigure}
    \caption{The distributions of the missing momentum. The left plot shows each HNL mass. The left plot shows the signal and background for a HNL mass of 100 GeV in the same format as \figref{fig:mvalep}.}
    \label{fig:pmis}
\end{figure}

The final rectangular cut is on the missing momentum. This is highly dependent on the HNL mass and a different cut value is therefore used for each mass point. The missing momentum for each mass point is shown in \figref{fig:pmis} (left). 
Both lower and upper bound cuts are used to reduce all types of backgrounds. The signal and background distributions can be seen in \figref{fig:pmis}, right plot. In this case, for a HNL mass of 100 GeV, the missing momentum is required to be between 10 and 45 GeV (cut 6).

As shown in \tabref{tab:cuttable} after all the rectangular cuts, the background is dominated by the irreducible $qqh$ background, followed by remaining 4-fermion hadronic and semi-leptonic background. The signal over background ratio ($S/B$) is already improved by 4 orders of magnitude, from $1/10^{5}$ in the beginning to $1/10$. The same cut variables are used for both beam polarizations but the cut values are optimized separately, as the background and signal differs for the two beam polarizations. Generally, the cuts for $(+0.8, -0.3)$ are looser, as the background is lower. Two cuts that were highly dependent on the HNL mass, specifically the first and last rectangular cuts, were optimized and tuned for each mass point. 

\subsection{Machine learning}
The signal and background events that passed the rectangular cuts were further filtered through a boosted decision tree (BDT) using the TMVA framework \cite{tmva}. One BDT was trained for each mass point and beam polarization. 
In total, 13 input parameters were passed to the BDT. The input parameters are listed below.
\begin{itemize}
    \item The lepton and missing energies
    \item 4-jet combined momentum
    \item The angle between the lepton and the closest jet
    \item $\cos\theta_l, \cos\theta_\nu, \cos\theta_Z, \cos\theta_{N_d}$, where $\theta$ is the production angle (in the lab frame) of the particle indicated by the subscript. The particles are the isolated lepton, reconstructed neutrino, Z boson and HNL, respectively
    \item The cosine of the lepton helicity angle in the HNL rest frame
    \item The reconstructed Higgs, $Z$ boson, and $W$ boson masses
    \item The corrected reconstructed HNL mass
\end{itemize}

\begin{figure}
    \centering
    \includegraphics[width=.6\linewidth,trim={0 0 0 1cm},clip]{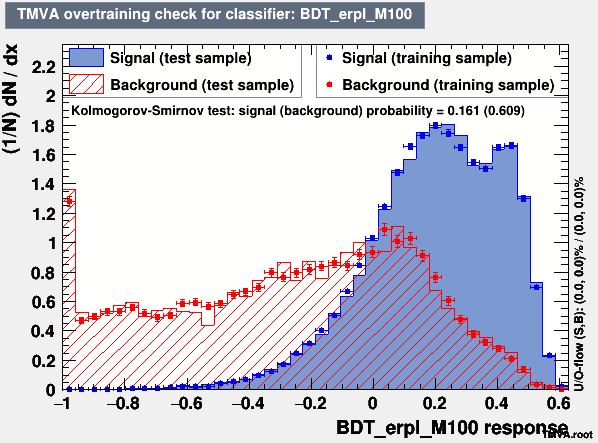}
    \caption{The distributions of BDT output for a BDT trained on the signal and background events when the HNL mass is 100 GeV and the beam polarization is $(+0.8, -0.3)$. The filled histograms show the signal and background events for the test data, while the dotted histograms show the training data.}
    \label{fig:overtrain}
\end{figure}

The formula for the corrected reconstructed HNL mass is $m(N_d) - m_W + m_{W_0}$, where $m_W$ is the reconstructed $W$ boson mass and $m_{W_0} = 80.4$ GeV is the truth central value of the $W$ boson mass. This corrected HNL mass has better resolution than that directly reconstructed from the lepton and two jets from $W$ since the largest uncertainty from the jet reconstruction can mostly cancel out. The angles are calculated by using the reconstructed particles. For the case of $\cos\theta_\nu$, the missing 4-momentum is used and assumed to be the neutrino momentum. 
Distributions of some example input parameters for the signal and background are shown in \figref{fig:sig_bkg_bdt}. The rest of the input parameters to the BDT can be found in the appendix. One example of the BDT output distributions is shown in \figref{fig:overtrain}, where it is validated that the BDT was not overfitted. Here, it can clearly be seen that the BDT output distribution is nearly identical for the training (dots) and test (filled) events. The Kolmogorov-Smikronov test also gives a high value (above 0.05) which quantitatively shows that the distributions are similar to each other.
A final cut is applied to the BDT output which further improves the $S/B$ by a factor of 3, as shown in \tabref{tab:cuttable}, and after the final cut the background contribution is completely dominated by $qqh$ process.

\begin{figure}
    \begin{subfigure}[t]{.49\linewidth}
        \centering
        \includegraphics[width=\linewidth]{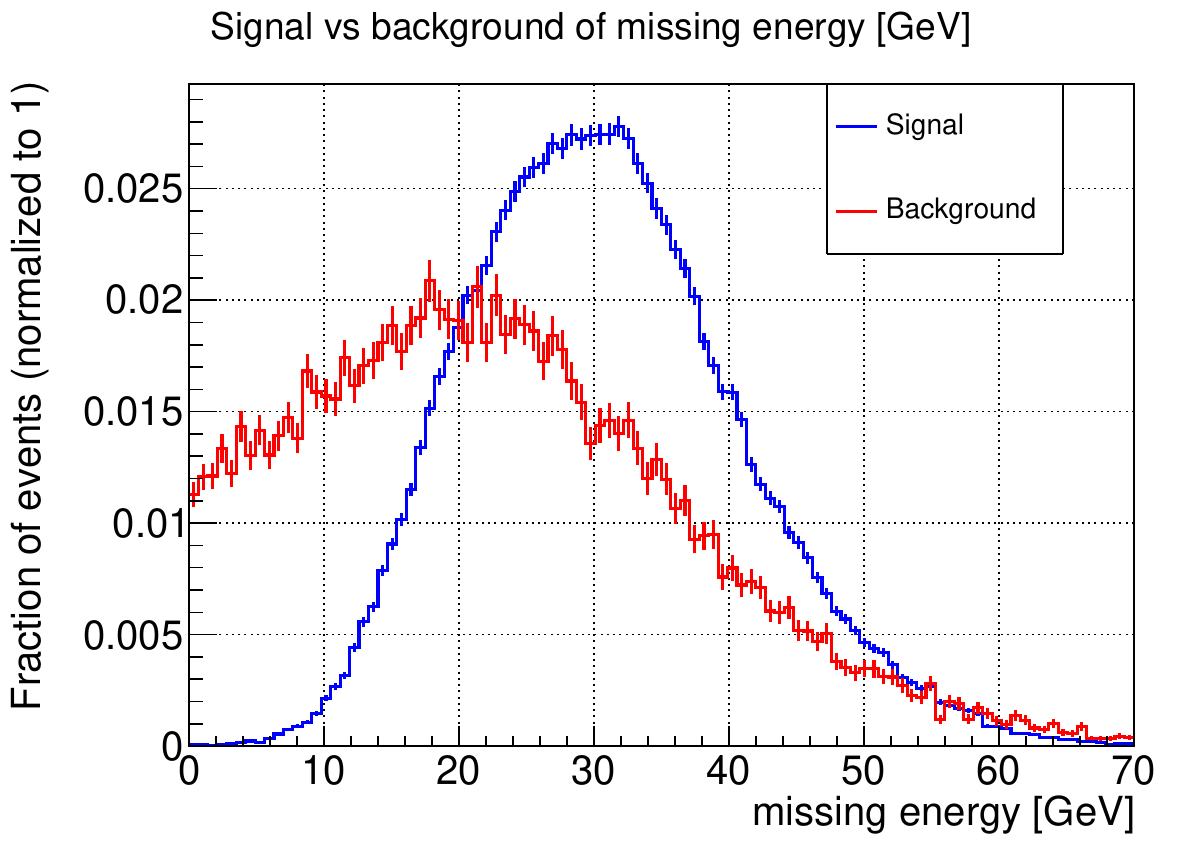}
    \end{subfigure}
    \hfill
    \begin{subfigure}[t]{.49\linewidth}
        \centering
        \includegraphics[width=\linewidth]{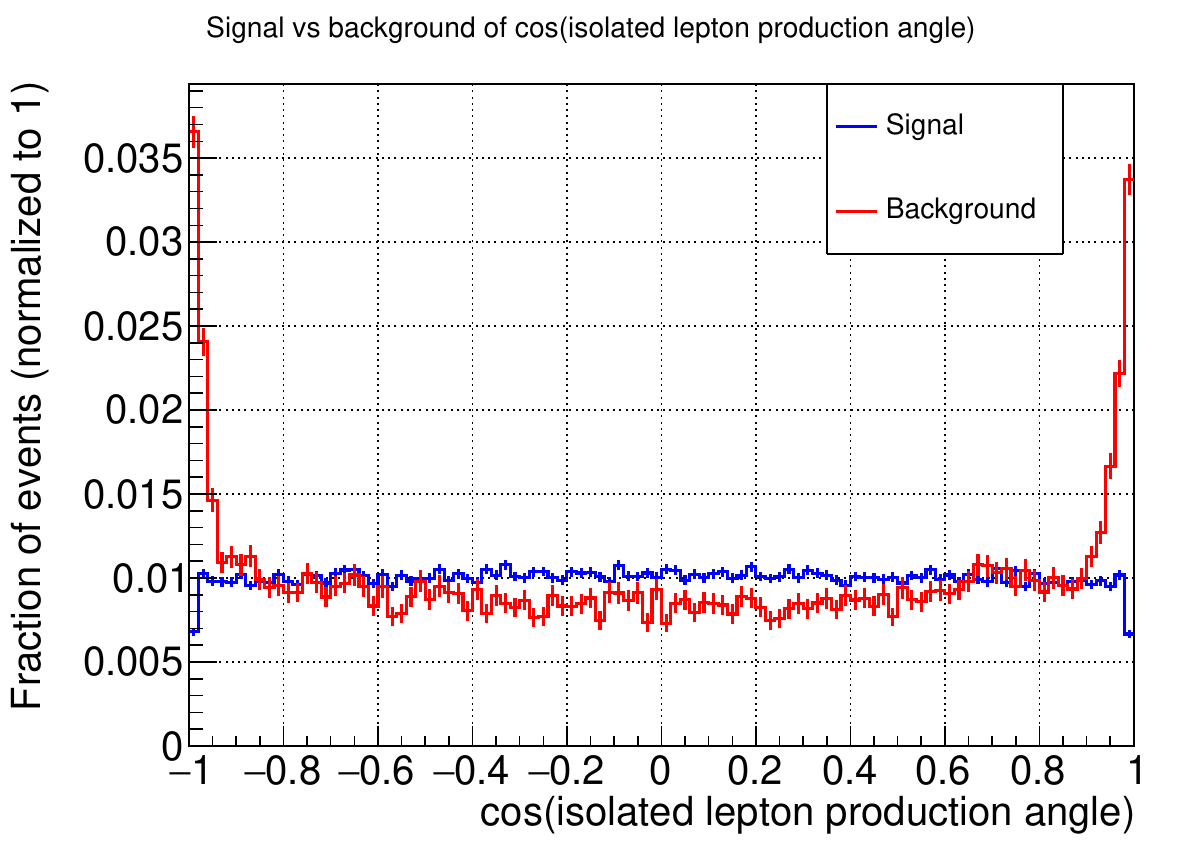}
    \end{subfigure}
    \hfill
    \begin{subfigure}[t]{.49\linewidth}
        \centering
        \includegraphics[width=\linewidth]{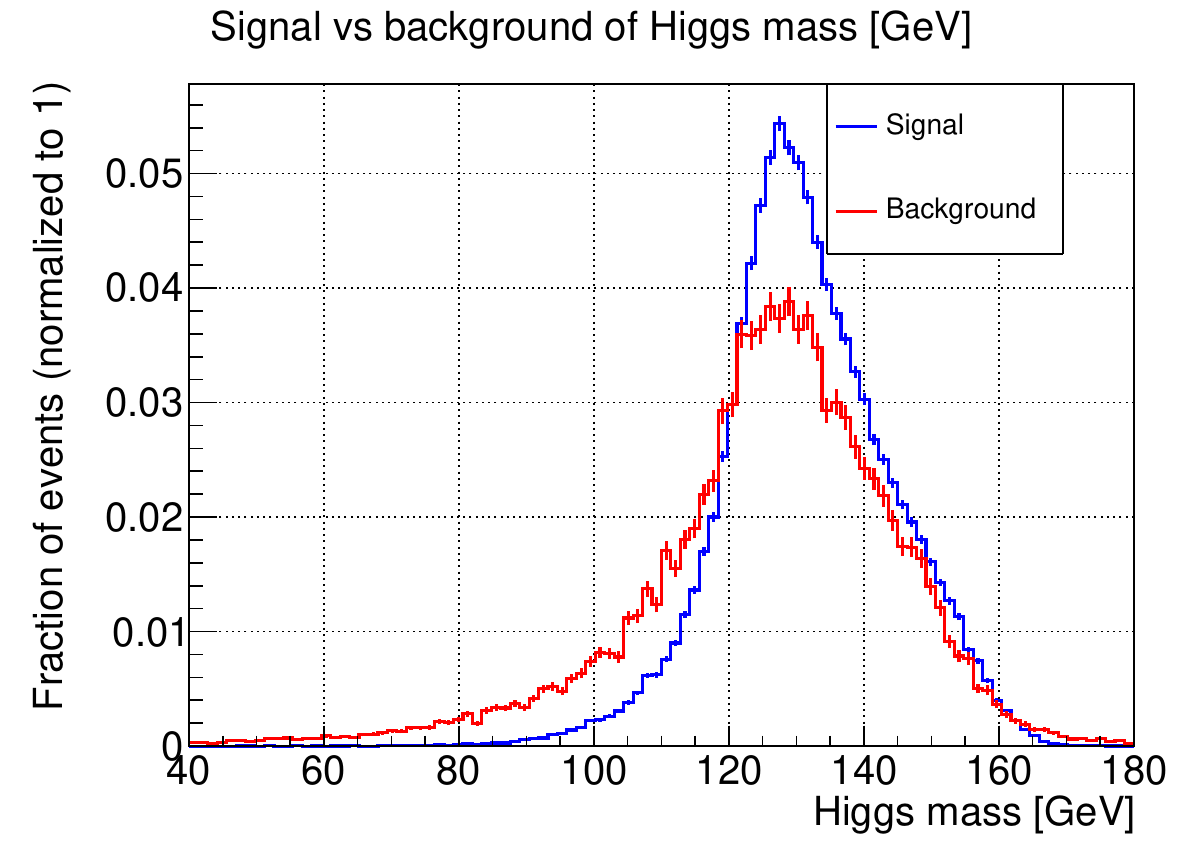}
    \end{subfigure}
    \hfill
    \begin{subfigure}[t]{.49\linewidth}
        \centering
        \includegraphics[width=\linewidth]{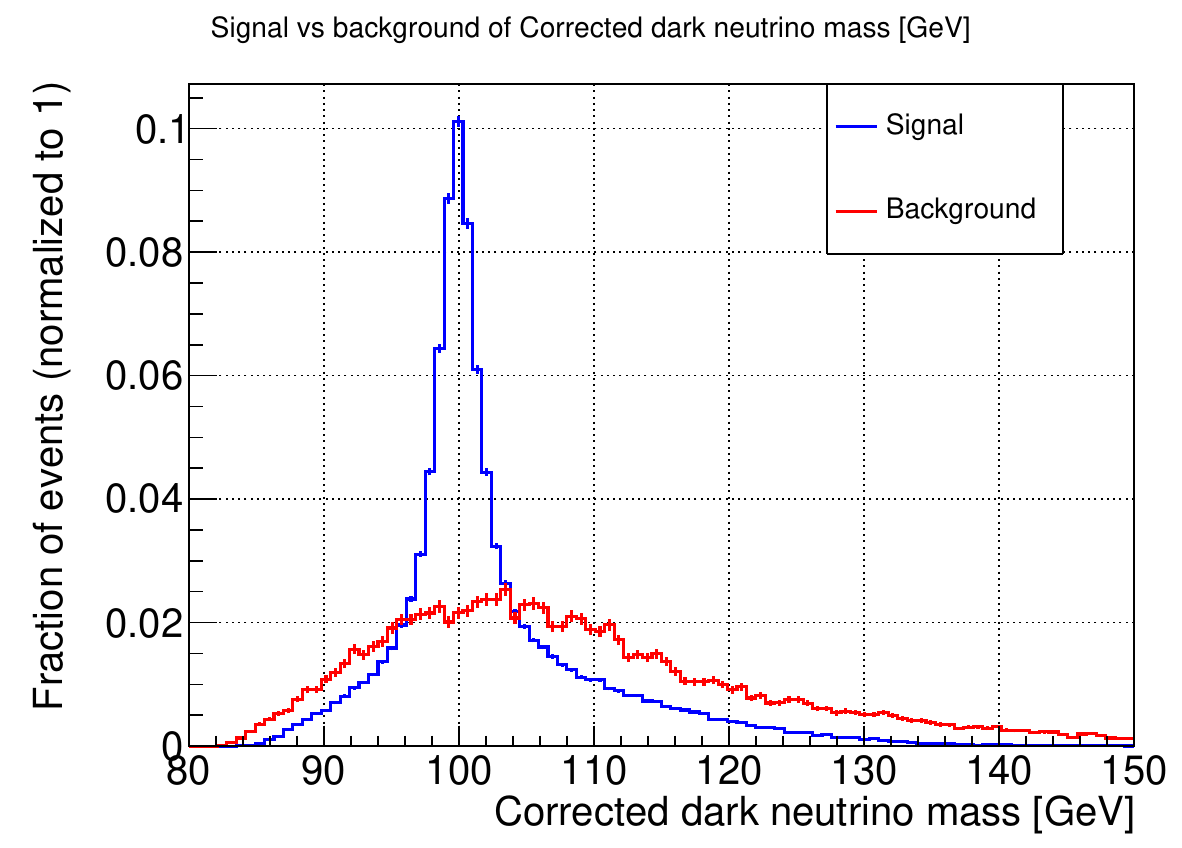}
    \end{subfigure}
    \caption{Distributions of a few input parameters to the BDT for the signal and background events, with the format as same as in e.g., \figref{fig:mvalep}, except that there is no grey region indicated. As in the previous figures, a HNL mass of 100 GeV and a beam polarization of $(+0.8, -0.3)$ is shown. From top left to bottom right with the first row first: missing energy, cosine of lepton production angle in the lab frame, the Higgs mass, and the corrected HNL mass.}
    \label{fig:sig_bkg_bdt}
\end{figure}

It is worth pointing out that by comparing the truth and reconstructed information we found that the jet clustering and jet pairing to distinguish jets from $Z$ and jets from $W$ is not perfect. Instead, jet constituents can originate from both $Z$ and $W$ bosons, which result in a bias in the jet momentum. This is discussed in further detail in the appendix.

\section{Results} \label{sec:exclusion}
\begin{figure}
    \centering
    \begin{subfigure}[t]{\linewidth}
        \centering
        \includegraphics[width=\linewidth,trim={0 87cm 42cm 0},clip]{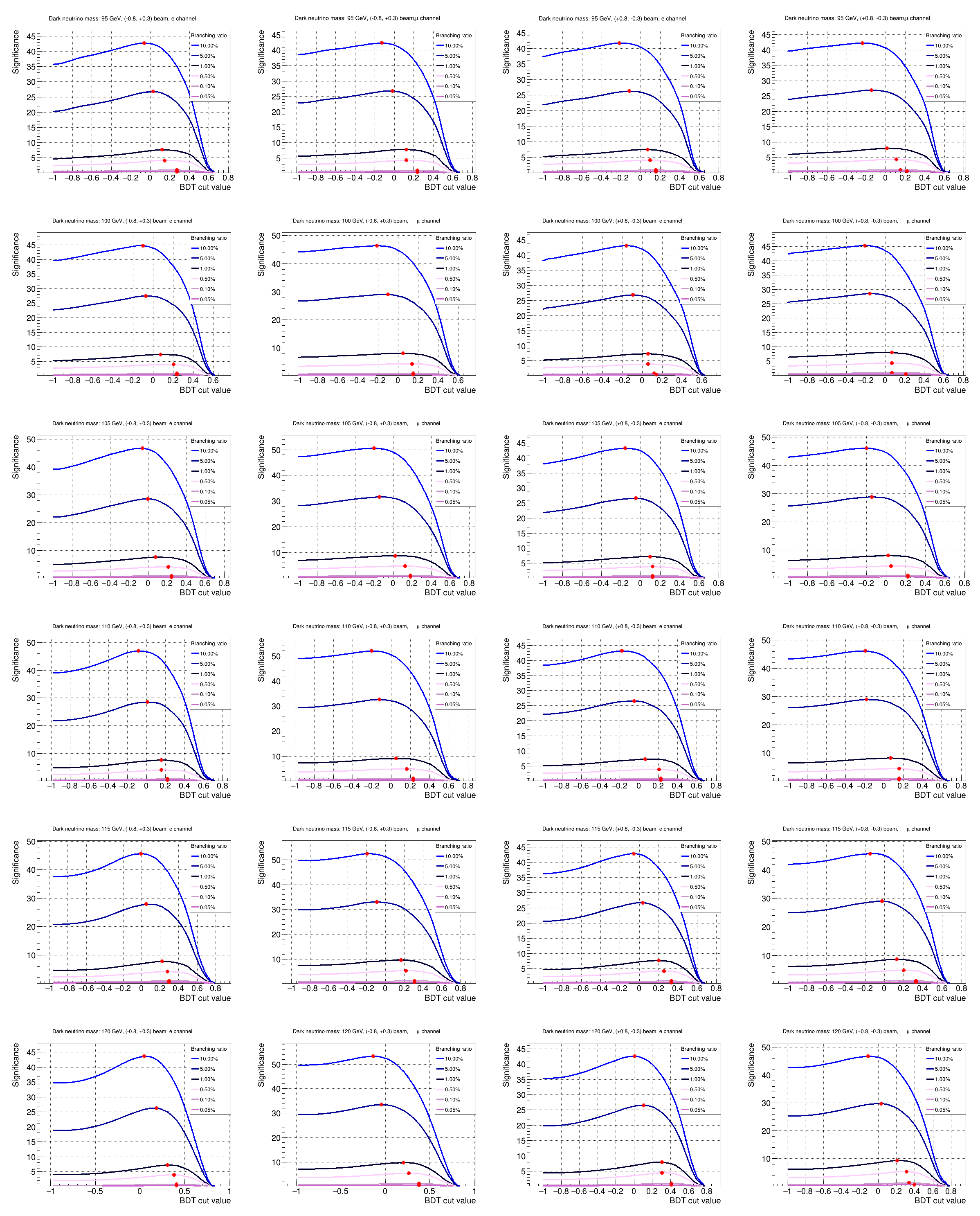}
    \end{subfigure}
    \hfill
    \begin{subfigure}[t]{\linewidth}
        \centering
    \includegraphics[width=\linewidth,trim={42cm 87cm 0 0},clip]{figures/significance_curve_BDT_mnd_m_br.png}
    \end{subfigure}
    \caption{Examples of significance curves as function of the BDT cut value for different values of branching ratio. All plots are for HNL masses of 95 GeV, with the top and bottom row being for $(-0.8, +0.3)$ and $(+0.8, -0.3)$ beam polarization respectively. The left (right) column of plots show the electron (muon) channel. Red dots indicate the location of maximum significance for different branching ratios where corresponding final cut is applied. The significance curves for other HNL masses and beam polarizations typically have a similar shape.
    }
    \label{fig:sig_curves}
\end{figure}

The analysis above illustrated mostly for one benchmark scenario, is carried out for each of the six values of HNL mass from 95 GeV to 120 GeV, each of the two beam polarization schemes and each of the two lepton channels ($e$ and $\mu$). The final BDT cut to maximize the signal significance also depends on the value of joint branching ratio as shown in \figref{fig:sig_curves} for a range of branching ratios from $0.01\%$ to $10\%$. The final significance for each branching ratio and HNL mass is then calculated by combining the contribution from two beam polarization schemes and two lepton channels. This combination is calculated as 
\begin{equation}
    \sigma_{final} = \sqrt{\sum_i \sigma_{i}^2}
\end{equation}
where $i$ iterates over the four ($2\cdot2$) possible combinations of lepton channels and beam polarizations.

\begin{figure}
    \centering
    \begin{subfigure}[t]{\linewidth}
        \centering
        \includegraphics[width=\linewidth,trim={0.2cm 21.5cm 8.9cm 0},clip]{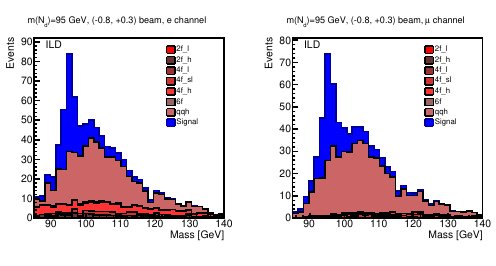}
    \end{subfigure}
    \hfill
    \begin{subfigure}[t]{\linewidth}
        \centering
        \includegraphics[width=\linewidth,trim={9cm 21.5cm 0.2cm 0},clip]{figures/mnd_adjusted_BDT_nomnd_m_postMVA.pdf}
    \end{subfigure}
    \caption{Examples of mass distributions after all cuts, with a branching ratio of $1\%$. The final BDT cut does not use the neutrino mass as input for this figure. The order of the plots is the same as in \figref{fig:sig_curves}. The backgrounds are categorized based on if a process result in 2 leptons (2f\_l), 2 quarks (2f\_h), 4 leptons (4f\_l), 2 leptons and 2 quarks (4f\_sl), 4 quarks (4f\_h), 6 fermions (6f), or 2 quarks and a Higgs boson (qqh). The signal is marked with blue.}
    \label{fig:mass_distributions}
\end{figure}

To better visualize the remained signal and background events, we separately trained a BDT without using the corrected HNL mass as one input variable, applied a cut on that BDT output, and then plotted the distributions of the corrected HNL mass, as shown in \figref{fig:mass_distributions}. If the branching ratio is as large as $1\%$, a sharp resonance from the HNL signal events would be clearly visible on top of the background events. 

\begin{figure}
    \centering
    \includegraphics[width=\linewidth]{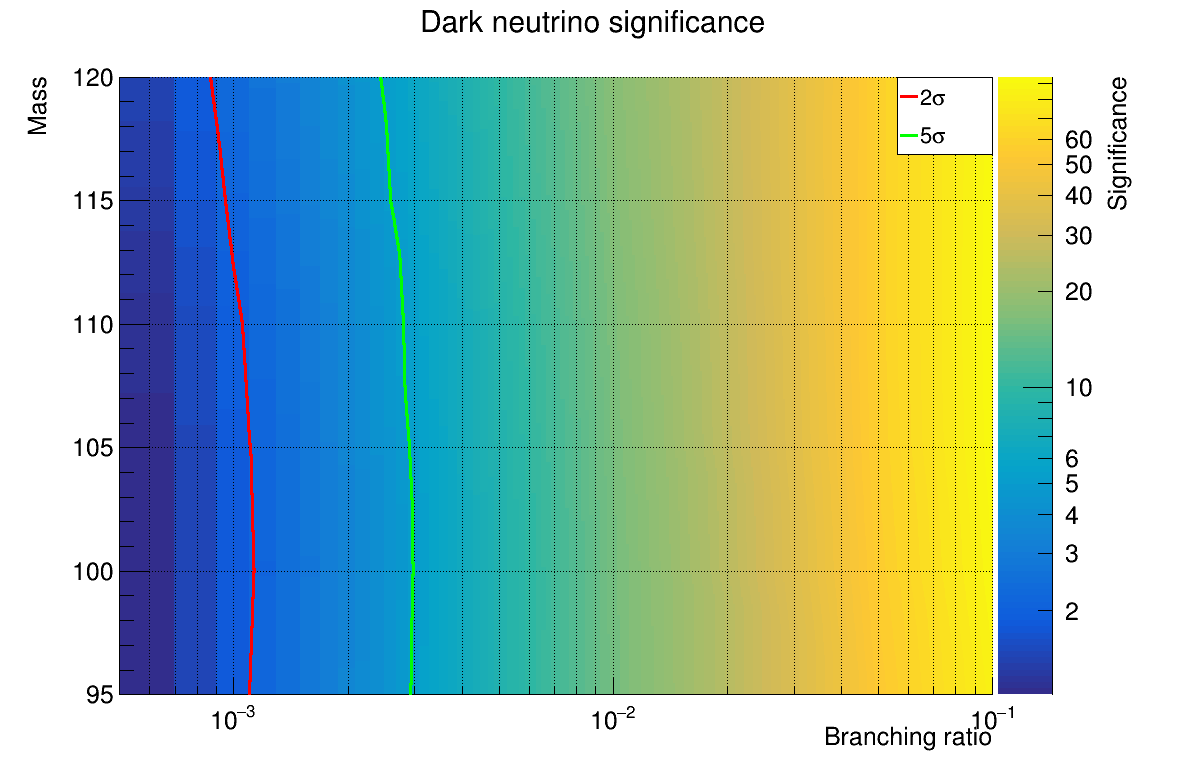}
    \caption{Exclusion plot as a function of branching ratio $BR(H\to \nu N_d)BR(N_d\to lW)$ ($x$ axis) and HNL mass ($y$ axis). The color indicates the significance of detecting this channel. The red curve indicates where the significance is $2\sigma$, i.e., the exclusion limit, whereas the green line shows the 5$\sigma$ limit, i.e., the limit for discovery. Every parameter value to the right of the red (green) line can be excluded (discovered). Note that a log scale is used on the $x$ and $z$ axis. The significance is interpolated in between the tested mass points and branching ratios and may not be completely accurate.}
    \label{fig:exclusion}
\end{figure}

We are now ready to give our model-independent final result in terms of the signal significance for searching the HNLs at the ILC250 as a function of the HNL mass and the joint branching ratio of $BR(H\to \bar\nu N_d)\cdot BR(N_d\to l W)$. The result is shown in \figref{fig:exclusion}, where the red line indicates the exclusion limit for $2\sigma$ significance and the green line indicates the discovery potential for $5\sigma$ significance. 
We can see in \figref{fig:exclusion} that the significance is almost identical regardless of HNL mass, with a slight decrease close to 100 GeV. This is most likely due to the background distribution, which has a peak close to 100-105 GeV (see \figref{fig:mass_distributions}). This means that it is more difficult for the BDT to discriminate the signal from the background, since the mass peaks overlap. As a consequence of this, the significance is highest at 120 GeV, since this is the farthest away from the background mass peak.
As a conclusion the exclusion limit (discovery potential) for the joint branching ratio is about $0.1\%~(0.3\%)$, nearly independent of the HNL mass given the mass range between $m_Z$ and $m_H$. For one study of the High Luminosity LHC (HL-LHC) with the same final state as the one investigated in this study, it was found that a branching ratio of $BR(H\to N_d \nu) = 1\%$ would have a $0.4\sigma$ signal significance \cite{hl-lhc-hn}. In comparison, for this study we expect the signal significance to be around $10\sigma$ (since $BR(N_d\to lW) \approx 80\%)$, which is an improvement of the significance of a factor of 25 for the ILC compared to the HL-LHC.

\begin{figure}
    \centering
    \begin{subfigure}[t]{.49\linewidth}
        \centering
        \includegraphics[width=\linewidth]{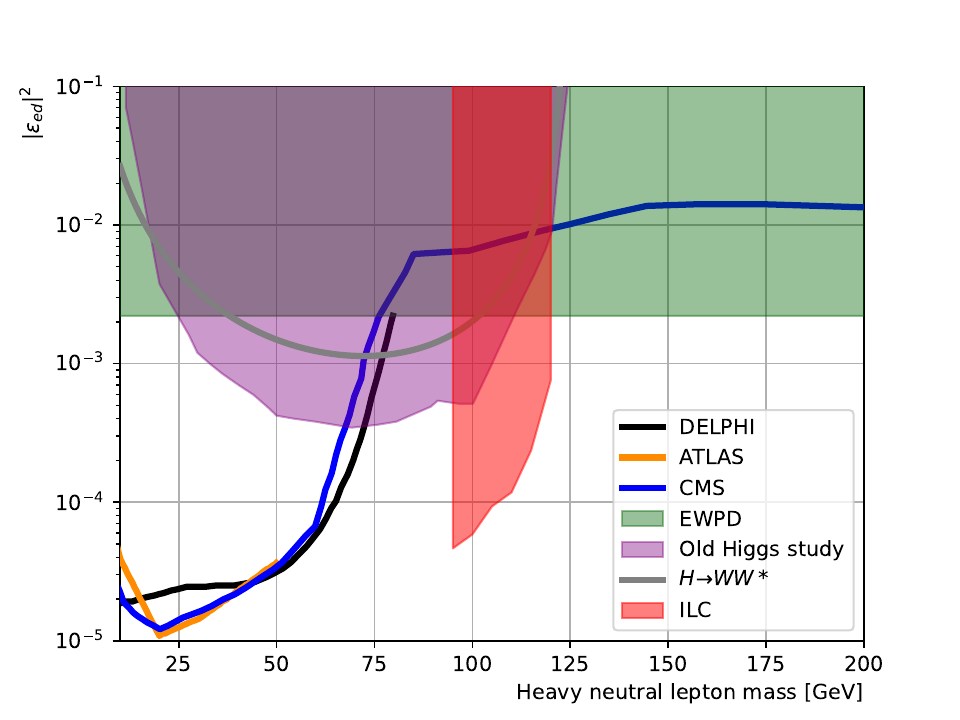}
    \end{subfigure}
    \hfill
    \begin{subfigure}[t]{.49\linewidth}
        \centering
        \includegraphics[width=\linewidth]{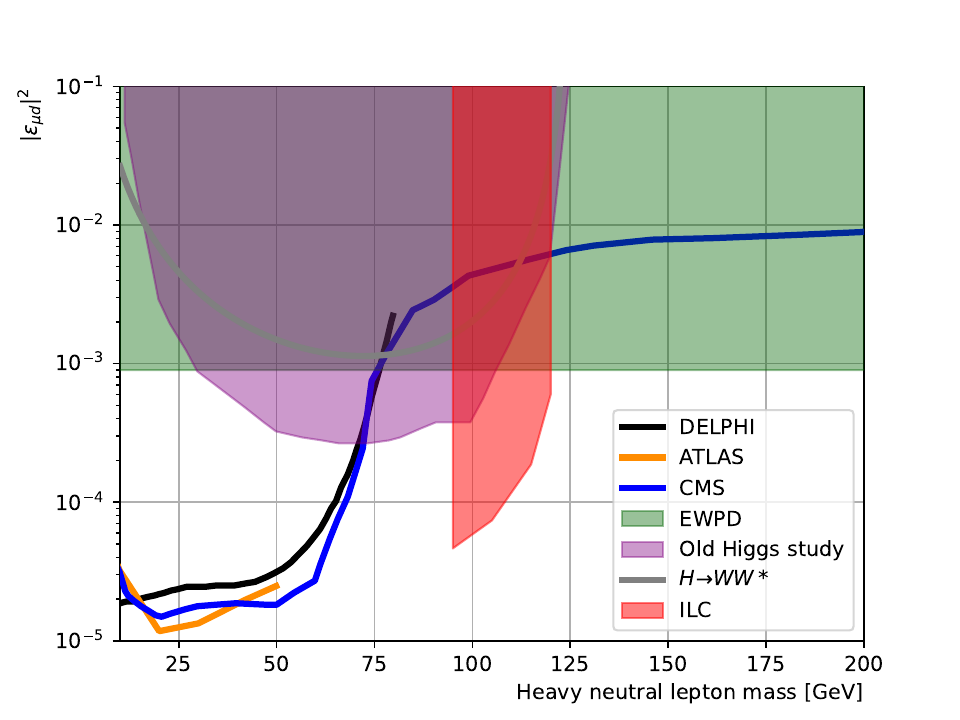}
    \end{subfigure}
    \caption{The exclusion curves for the mixing matrix elements between the HNL and the electron neutrino (left plot) and the muon neutrino (right plot). The $x$-axis is the HNL mass while the $y$-axis is the squared amplitude of the mixing matrix element. Different constraints have different colors, with the black line showing searches by DELPHI at LEP, the orange line showing the ATLAS trilepton search, the blue line corresponding to the CMS trilepton search, the green region corresponding to measurements of the SM neutrino mixing matrix elements (i.e., electroweak precision data), the purple region showing the results from \cite{nd_higgs}, and the gray line showing the region where the $BR(H\to \nu N_d) > 0.05$.
    The red region shows the new results from this study using simulated ILC data. Note that the $y$-axis is a logarithmic scale. The contours are linearly interpolated between each simulated mass point (for this study, it is 95 - 120 GeV, 5 GeV apart).
    }
    \label{fig:overlay_exclusion}
\end{figure}

The model-independent results can be cast into constraints on the two free model parameters, HNL mass $m_N$ and mixing parameter $|\varepsilon_{id}|^2$, as introduced in Section 1, using Equation \ref{eq:br2eps}. Note that in this interpretation, the branching ratio $BR(H\to \nu \bar N_d + \bar\nu N_d) = 2 BR(H\to \nu \bar N_d)$ is used, i.e., a factor of 2 is multiplied to the branching ratio in Equation \ref{eq:br2eps}. The limits are calculated separately for the electron and muon channels. The constraints from this study, together with constraints imposed by previous studies mentioned above are shown in \figref{fig:overlay_exclusion}. The exclusion limits from current constraints are taken from \cite{snc_paper, snc_website, atlas_trilepton}.
We can see that the exclusion limit on $|\varepsilon_{id}|^2$ down to $10^{-4}$ can be reached at the ILC250, in the HNL mass region between $Z$ mass and Higgs mass, which is about 1-order of magnitude improvement over the current constraint. Note however that the current constraint from Ref. \cite{nd_higgs} is done using fast simulations and assumes that the HNL only mixes with one SM neutrino. These constraints are therefore likely optimistic and the constraint from this study is likely much more than a factor 10 improvement. 

\subsection{Discussion for potential improvement}


As mentioned earlier one source of error in this analysis is the jet clustering and jet pairing. This was e.g., evident in the distribution of the lepton helicity angle in the HNL rest frame (see e.g., \figref{fig:l_helicity_mc_rec}). Improving this by using better jet clustering algorithms and/or a more sophisticated method for jet pairing could improve the measurements of the 4-momenta of the jets, which in turn improves the precision of the $W$ and $Z$ bosons reconstruction, and consequently of the HNL and Higgs boson reconstruction. It is therefore potentially of great help to improve jet clustering algorithms for future experiments and analyses that focus on precision measurements.

Another potential improvement to this study is on the correction of the HNL mass. Currently, a simple correction was used by subtracting the $W$ boson mass and adding a constant, which evidently worked well for this study, since the $W$ boson 4-momentum measurement was the main source of error. However, one could also employ techniques such as kinematic fitting to correct for errors more accurately in the $W$ boson 4-momentum, which could improve mass resolution of the HNL mass even more. This is also closely related to the jet clustering error mentioned above.

\begin{figure}
    \centering
    \begin{subfigure}{.49\linewidth}
        \includegraphics[width=\linewidth]{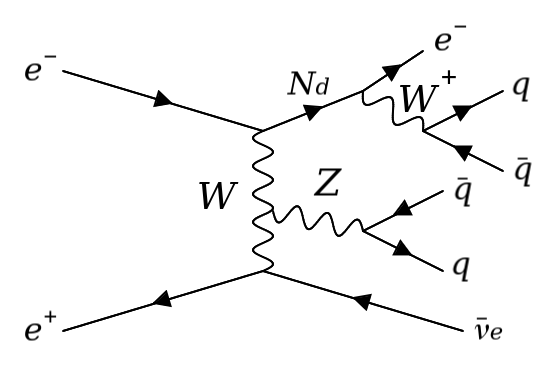}
        \caption{}
        \label{fig:hnl_background_a}
    \end{subfigure}
    \hfill
    \begin{subfigure}{.49\linewidth}
        \includegraphics[width=\linewidth]{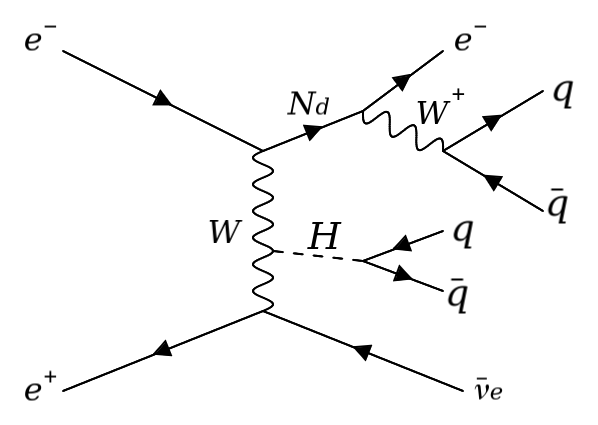}
        \caption{}
        \label{fig:hnl_background_b}
    \end{subfigure}
    \caption{Two Feynman diagrams with the same final state particles as the signal. The antiparticle version of these diagrams are also possible, where $e^+$ becomes $\bar N_d$ and $e^-$ becomes $\nu_e$.}
    \label{fig:hnl_background}
\end{figure}

There are two HNL decay channels that might affect the outcome of this study, which are shown in \figref{fig:hnl_background}. These two channels have the same final state as the main final state investigated in this study, and could therefore contribute to the number of events detected. If the contributions of these channels were significant, then $BR(N_d \to lW)$ would be constrained rather than $BR(H\to \nu N_d)\cdot BR(N_d\to lW)$. 
However, we expect the contributions from these diagrams to be relatively small, as many of them would be filtered by the BDT. 
Since the Feynman diagram in \figref{fig:hnl_background_a} has no Higgs boson, we expect that the cut that the BDT applies on the Higgs boson invariant mass will remove large portions of this channel. For the Feynman diagram in \figref{fig:hnl_background_b}, there is no Z boson, which means that the BDT cut on the Z boson mass would remove large portions of this channel. Additionally, the Higgs boson has other decay products than in our signal diagram, which means that the reconstructed Higgs invariant mass would be non-physical and deviate from 125 GeV significantly. This means that even more of this channel would be filtered and removed by the BDT.

To further improve the sensitivity to the HNL, other decay channels can be investigated. This includes the $\tau$ decay channel (i.e., $H\to \nu_\tau N_d \to \nu_\tau \tau q\bar{q}$) as well as two other $W$ and $Z$ boson decay channels. However, the other two decay channels are likely to provide little sensitivity. The semi-leptonic decay, i.e., a leptonic decay of the $W$ or $Z$ boson and a hadronic decay of the other one, would likely be difficult to discriminate from the background. This is because the final state contains two neutrinos which makes it difficult to reconstruct W and/or the Z boson and thus the HNL and Higgs. Additionally, there are two jets in the final state which has a large background. 
The other fully leptonic decay with three neutrinos and three leptons will likely have a very small background. However, reconstructing the HNL or any of the $W$, $Z$, and $H$ bosons will be almost impossible. Still, this decay channel was more sensitive than the hadronic decay channel for the LHC \cite{nd_higgs}, though this does not mean that it will be the same for ILC, since the background characteristics are different. It is therefore totally possible that this and other decay channels would have sizable contributions to the significance.

\section{Summary}
The sensitivity of the ILC for detecting heavy HNLs through exotic Higgs decays was investigated. The $e^+ e^- \to ZH \to q\bar{q}~ \nu N_d \to q\bar{q}~\nu~lW \to q\bar{q}~\nu l~q\bar{q}$ channel was studied, for the first time to our knowledge based on full detector simulations. The analysis was performed at 250 GeV center-of-mass energy for two different beam polarization schemes, and for six HNL masses between $m_Z$ and $m_H$. The SM background with 2, 4, 6-fermion final states as well as $q\bar{q}H$ processes were all considered in the analysis. Events were filtered through a pre-selection, a set of rectangular cuts, and finally a machine learning cut. The cuts were separately optimized for all combinations of beam polarizations and HNL masses, though the pre-selection was the same for all of them. The background contribution turns out to be very important, and only after all the dedicated event selection cuts the signal over background ratio could be improved from $O(1/10^{5})$ to $O(1)$. 
For all masses simulated, the dominating background after all cuts was from the SM Higgs decay process $H\to WW^* \to q\bar{q} l\nu$. The final significance achieved was around $2\sigma$ for a branching ratio of $BR(H\to\nu N_d)\cdot BR(N_d\to lW) = 0.1\%$, while $5\sigma$ is reachable at a branching ratio of $0.3\%$. The significance is almost unchanged for different masses. For a branching ratio of $BR(H\to\nu N_d)=1\%$, the significance is roughly $10\sigma$, which is 25 times greater than the expected performance of HL-LHC. Interpreting these results for HNL models results in constraints on the mixing matrix elements $|\varepsilon_{id}|^2$ between SM neutrinos and the HNL of levels down to $10^{-4}$, which is at least a factor of 10 improvement compared to previous constraints.


\section*{Acknowledgement}
We would like to thank Teresa N{\'u}{\~n}ez, Alberto Ru{\'i}z and Kiyotomo Kawagoe for constructive discussions and suggestions during the ILD internal review process. We also would like to thank Hitoshi Murayama and Robert McGehee for providing helpful theory guidance, ILD Monte Carlo Team in particular Hiroaki Ono and Ryo Yonamine for producing the common SM background samples, Krzysztof M{\k{e}}ka{\l}a and J\"urgen Reuter for pointing out the proper UFO model file for Whizard to generate signal events, and Kay Hidaka, Daniel Jeans, Shigeki Matsumoto, Satoshi Shirai, Taikan Suehara for discussions at various local meetings. This research was supported by the Sweden-Japan foundation and ``Insamlingsstiftelsen f\"or internationellt studentutbyte vid KTH''.

\bibliography{references}

\section{Appendix}
\subsection{Investigation of error in angles}
\begin{figure}
    \centering
    \includegraphics[width=.6\linewidth]{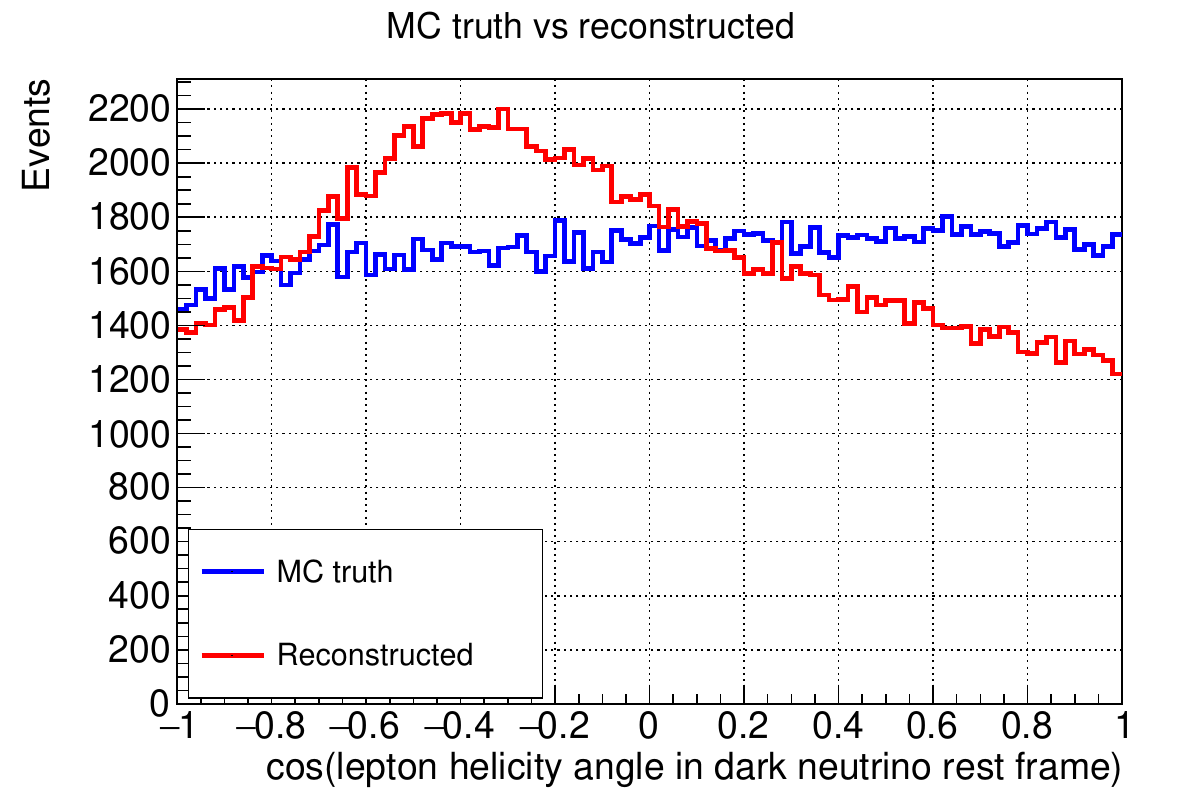}
    \caption{Distributions of the cosine of the lepton helicity angle in the HNL rest frame. The blue (red) histogram shows the MC truth (reconstructed) angular distribution. The $y$-axis shows the number of simulated events, without taking any cross section into account and is therefore arbitrary. The data shown is for a HNL mass of 110 GeV and a beam polarization of $(-0.8, +0.3)$ but the same pattern can be seen for all other signal events.}
    \label{fig:l_helicity_mc_rec}
\end{figure}

\begin{figure}
    \centering
    \includegraphics[width=\linewidth]{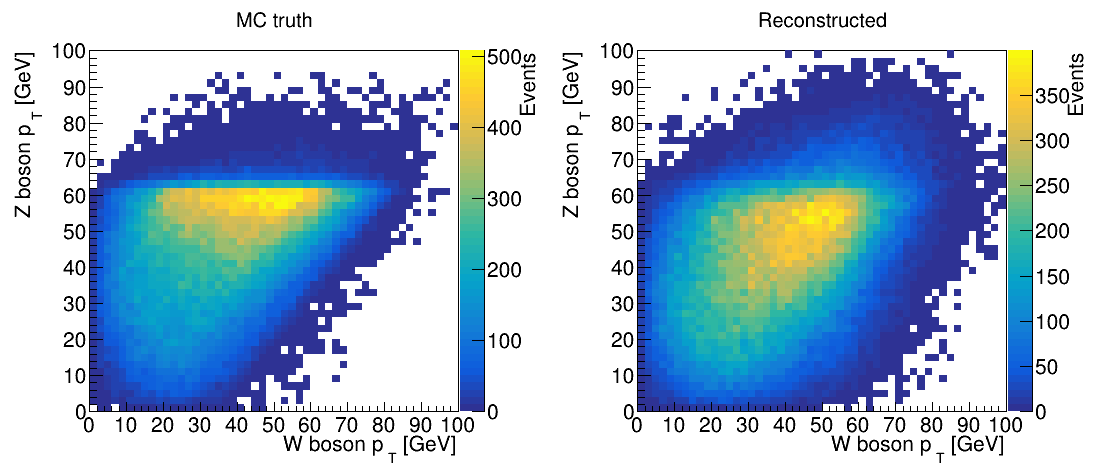}
    \caption{2D distributions of the W boson $\pt$ ($x$-axis) vs Z boson $\pt$ ($y$-axis). The left plot shows the MC truth distribution, while the right plot shows the reconstructed values. One can see that the concentrated cluster in the MC truth has been more spread out in the reconstructed version, with more values of higher W boson $\pt$ and either lower or higher Z boson $\pt$.}
    \label{fig:mc_rec_ptw_ptz}
\end{figure}

\begin{figure}
    \centering
    \includegraphics[width=\linewidth]{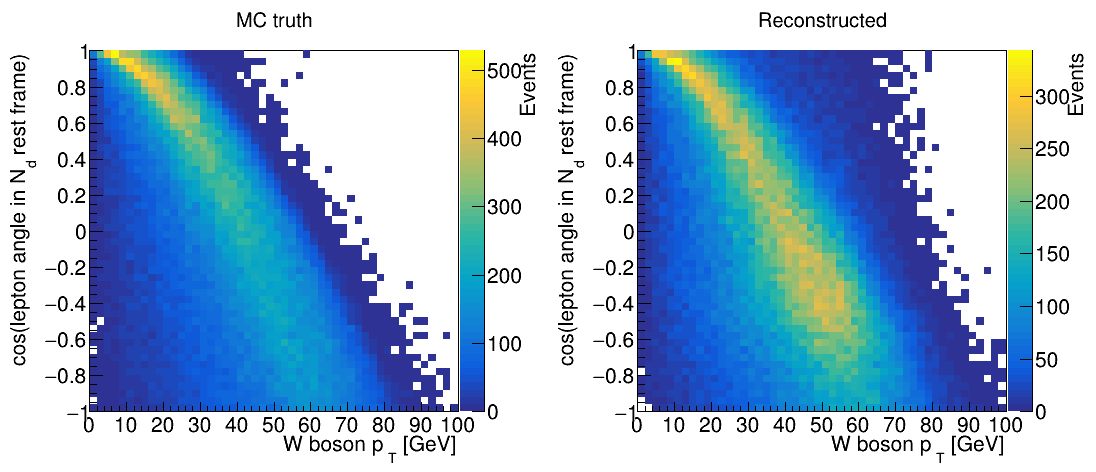}
    \caption{2D distributions of the W boson $\pt$ ($x$-axis) vs lepton helicity angle ($y$-axis). The format is the same as in \figref{fig:mc_rec_ptw_ptz}. One can see that the number of high-$\pt$ events have increased for reconstructed events and that these events typically appear for negative angles.}
    \label{fig:l_helicity_vs_ptw}
\end{figure}

One problem encountered when selecting input parameters for the BDT was with the lepton angle distribution in the HNL rest frame. When comparing the MC truth distribution with the reconstructed one, there is a slight shift towards negative angles for reconstructed angles (see \figref{fig:l_helicity_mc_rec}). This can be explained by incorrect jet pairing and errors in the jet clustering. 
Since the Z boson is more massive than the W boson, in general there will be more energy in the Z boson jets than the W boson ones. All jets are however soft due to the low center-of-mass energy and since the produced particles (Higgs, HNL etc) have a high mass. This makes it difficult for the jet clustering algorithm to correctly determine which particle should belong to which jet, and many particles originating from the Z boson can be misclassified as belonging to W bosons and vice versa. Furthermore, when performing the jet pairing, it is also possible for jets to incorrectly be assigned to the wrong boson, which further increases this error. As a consequence of this, the reconstructed W boson will sometimes have slightly higher energies and/or $\pt$ than MC truth. This is shown in \figref{fig:mc_rec_ptw_ptz}. We also see that the W bosons with high $\pt$ are the ones that contribute to an excess in negative helicity angles, as shown in \figref{fig:l_helicity_vs_ptw}. This therefore further supports our theory. To increase the accuracy of future analyses and collider experiments, improved jet clustering algorithms are therefore important.

\subsection{Additional figures}
All 1D parameter distributions of the input parameters to the BDT are shown below. This only shows the distributions for a signal with a HNL mass of 100 GeV and a beam polarization of $(+0.8, -0.3)$. Note that the $y$-axis has been normalized to only include the number of events. In reality, the signal histogram is much smaller relative to the background.

\begin{figure}
    \begin{subfigure}[t]{.3\linewidth}
        \centering
        \includegraphics[width=\linewidth]{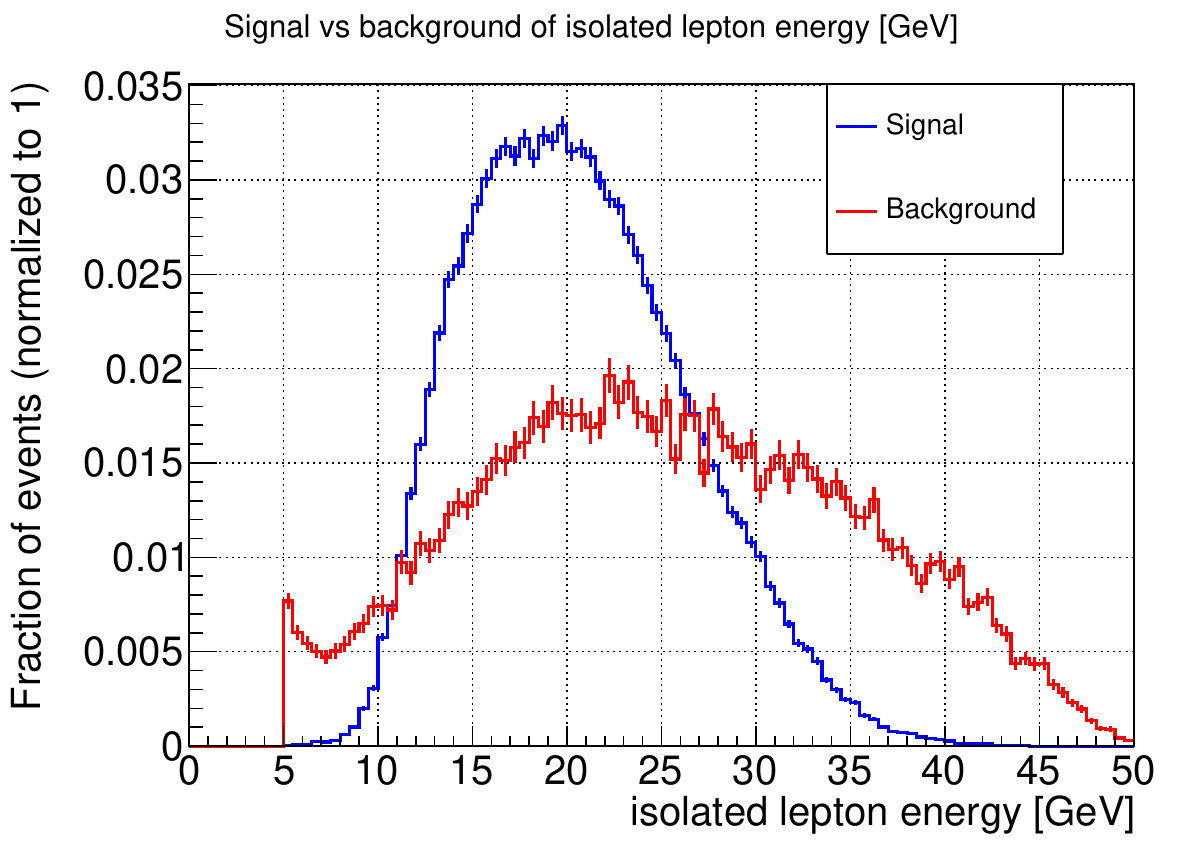}
    \end{subfigure}
    \hfill
    \begin{subfigure}[t]{.3\linewidth}
        \centering
        \includegraphics[width=\linewidth]{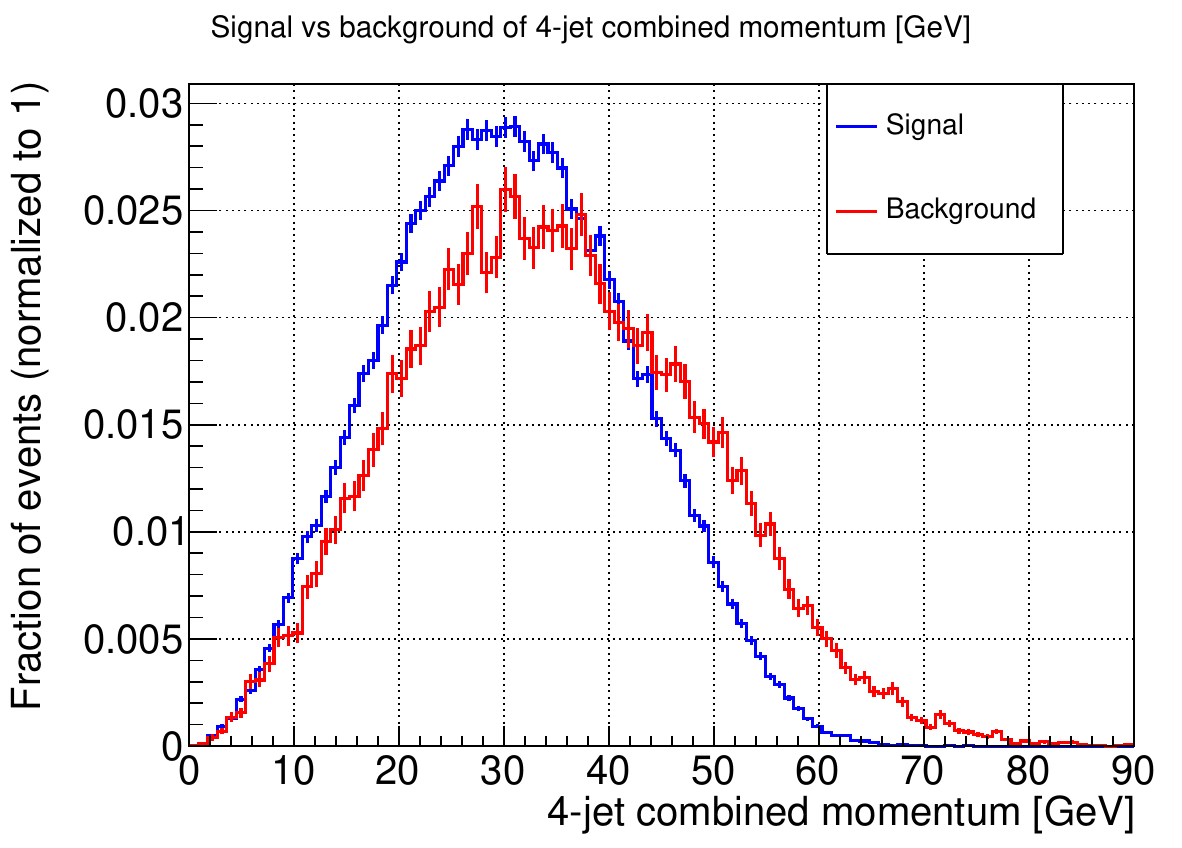}
    \end{subfigure}
    \hfill
    \begin{subfigure}[t]{.3\linewidth}
        \centering
        \includegraphics[width=\linewidth]{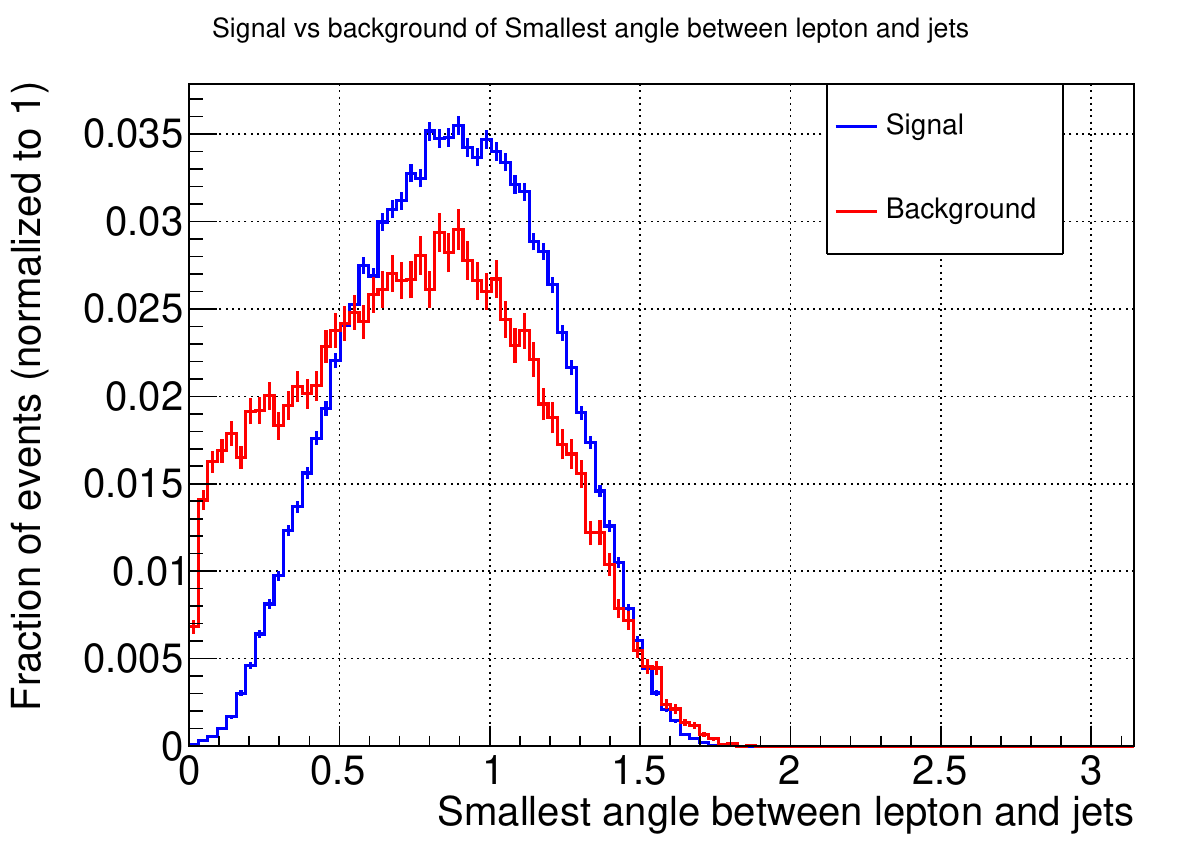}
    \end{subfigure}
    \hfill
    \begin{subfigure}[t]{.3\linewidth}
        \centering
        \includegraphics[width=\linewidth]{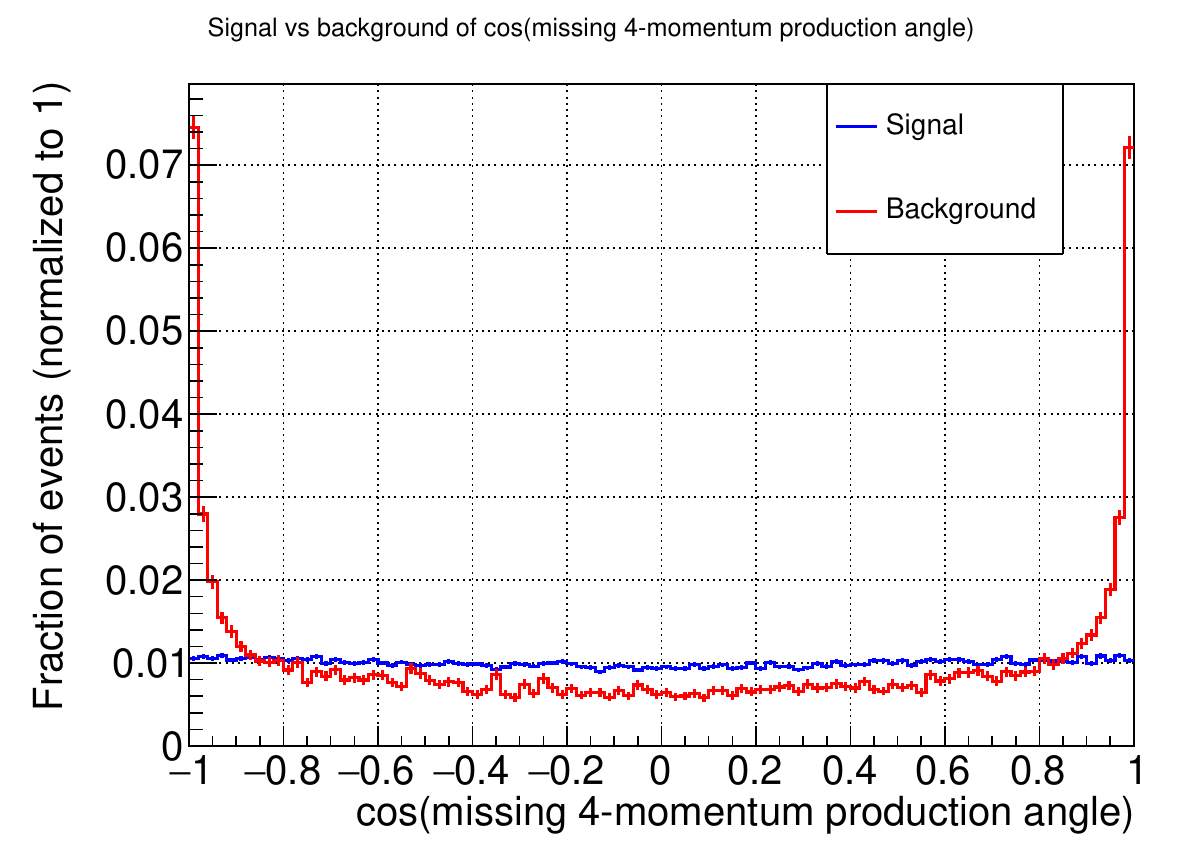}
    \end{subfigure}
    \hfill
    \begin{subfigure}[t]{.3\linewidth}
        \centering
        \includegraphics[width=\linewidth]{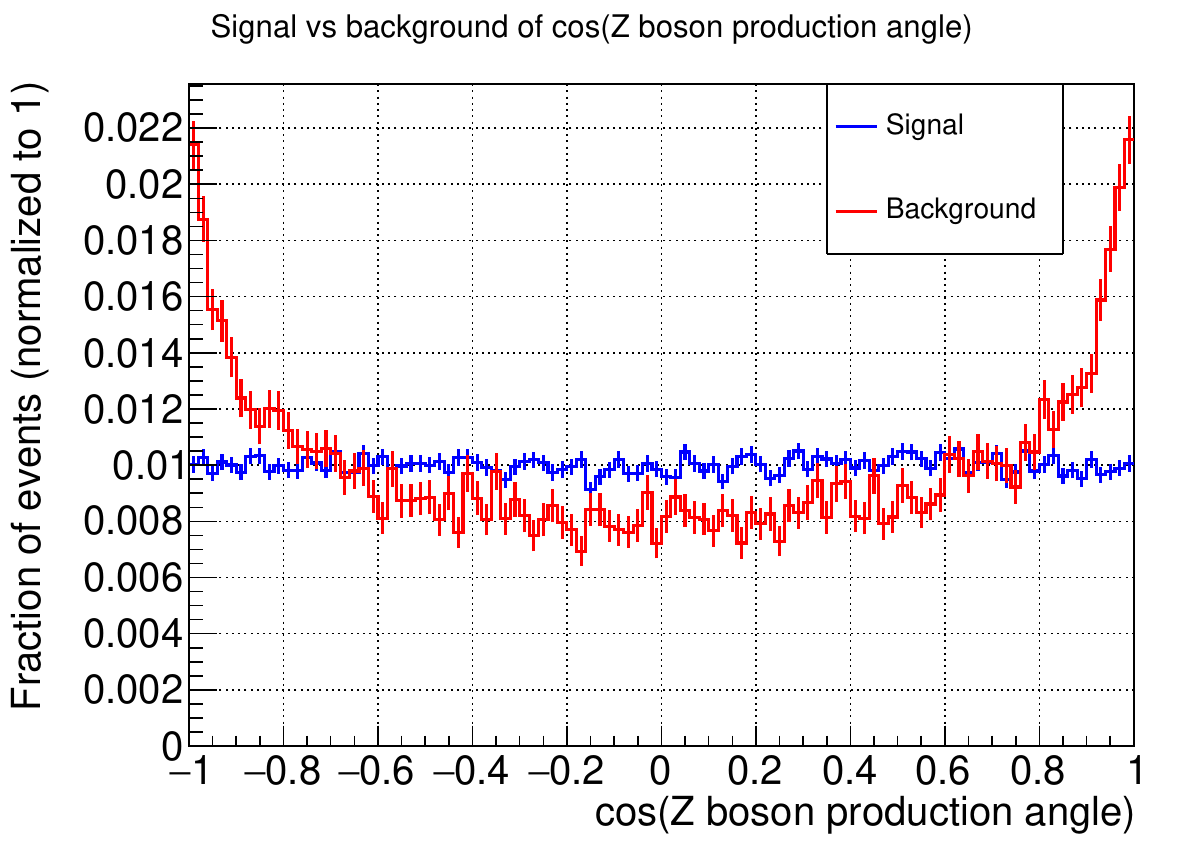}
    \end{subfigure}
    \hfill
    \begin{subfigure}[t]{.3\linewidth}
        \centering
        \includegraphics[width=\linewidth]{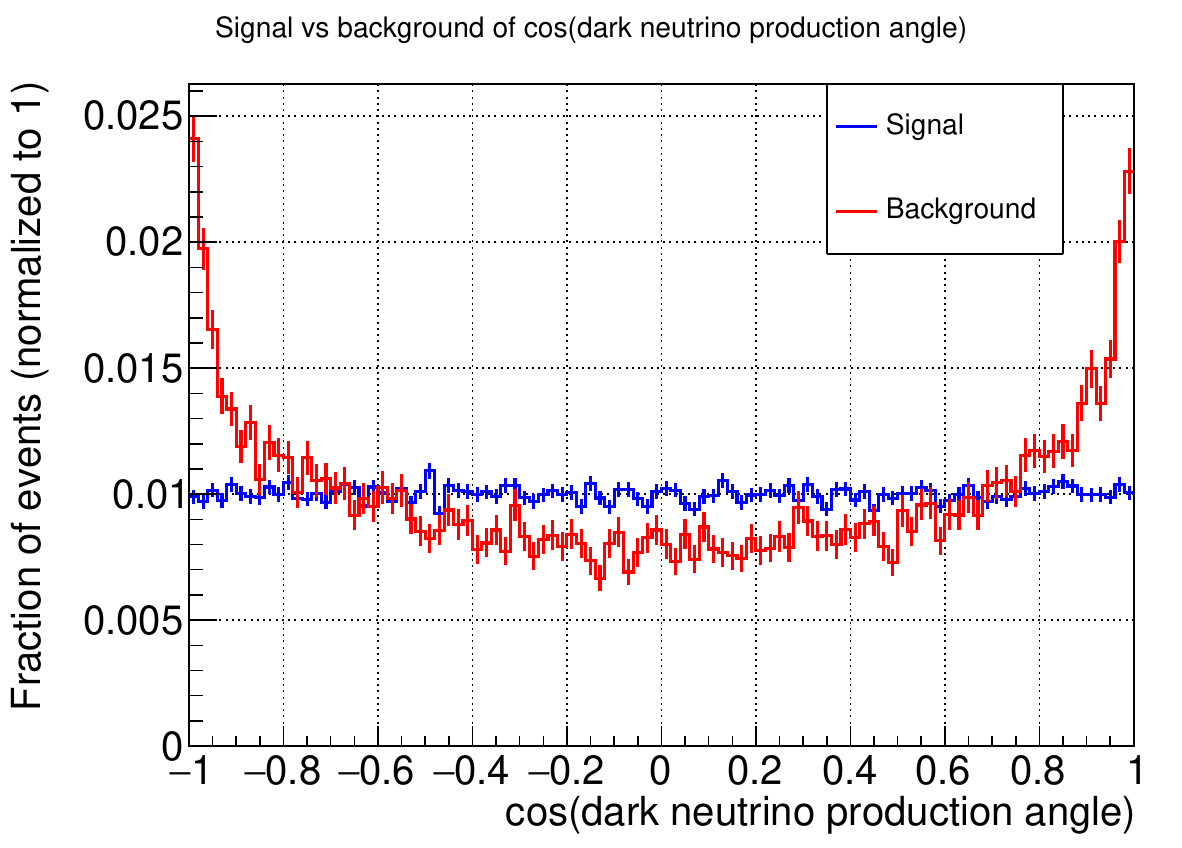}
    \end{subfigure}
    \hfill
    \begin{subfigure}[t]{.3\linewidth}
        \centering
        \includegraphics[width=\linewidth]{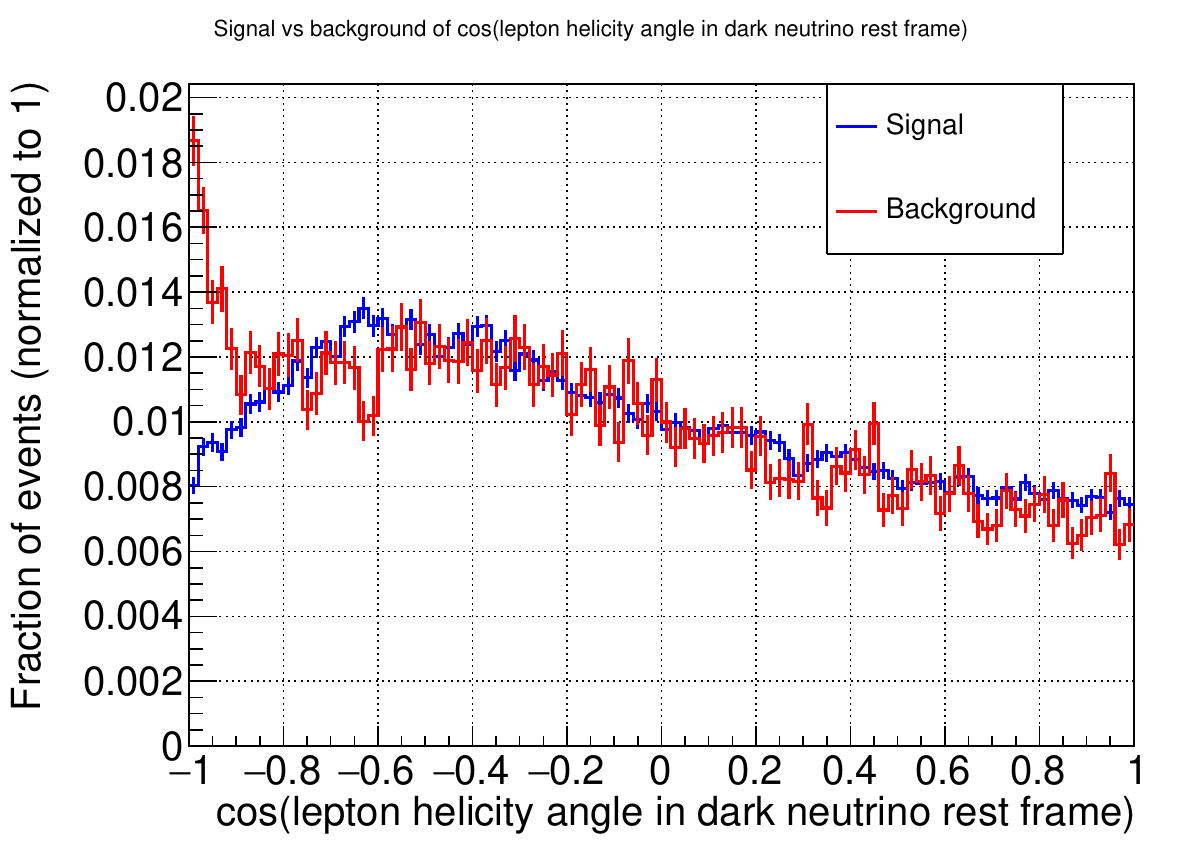}
    \end{subfigure}
    \hfill
    \begin{subfigure}[t]{.3\linewidth}
        \centering
        \includegraphics[width=\linewidth]{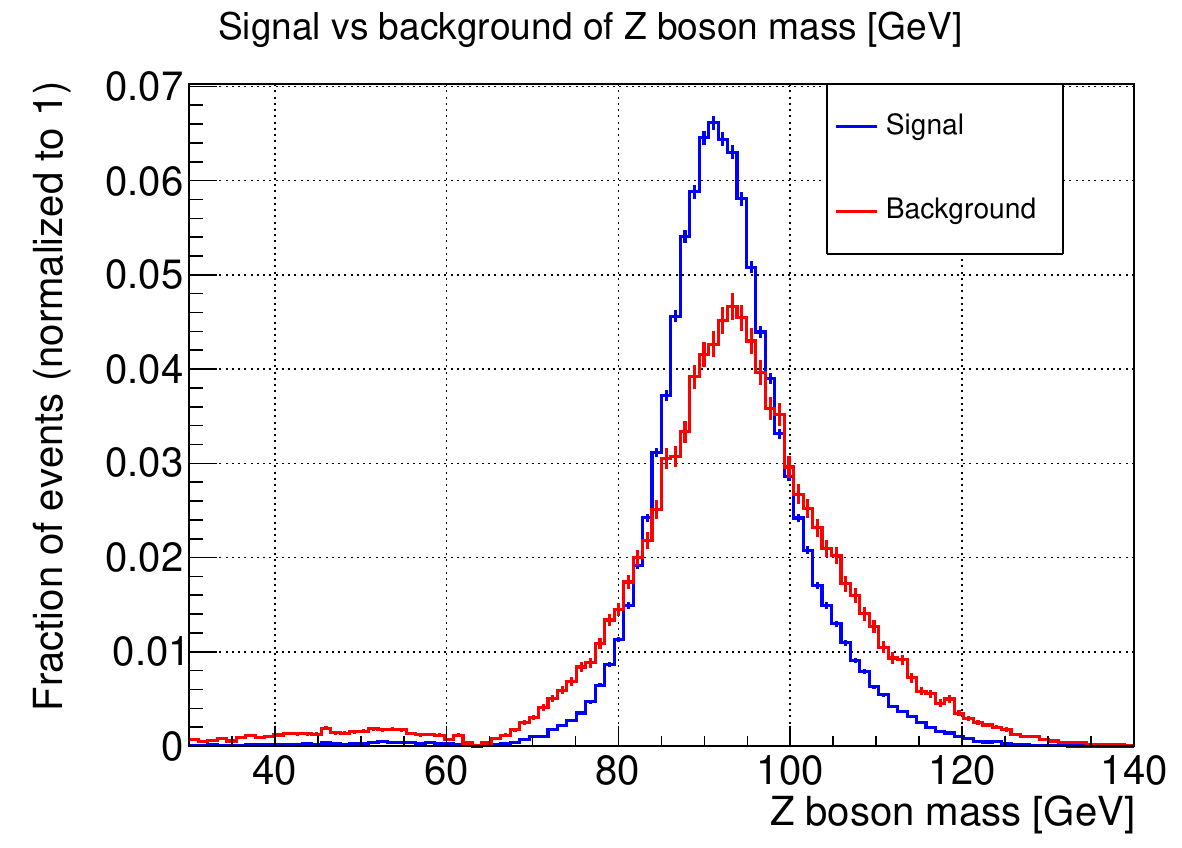}
    \end{subfigure}
    \hfill
    \begin{subfigure}[t]{.3\linewidth}
        \centering
        \includegraphics[width=\linewidth]{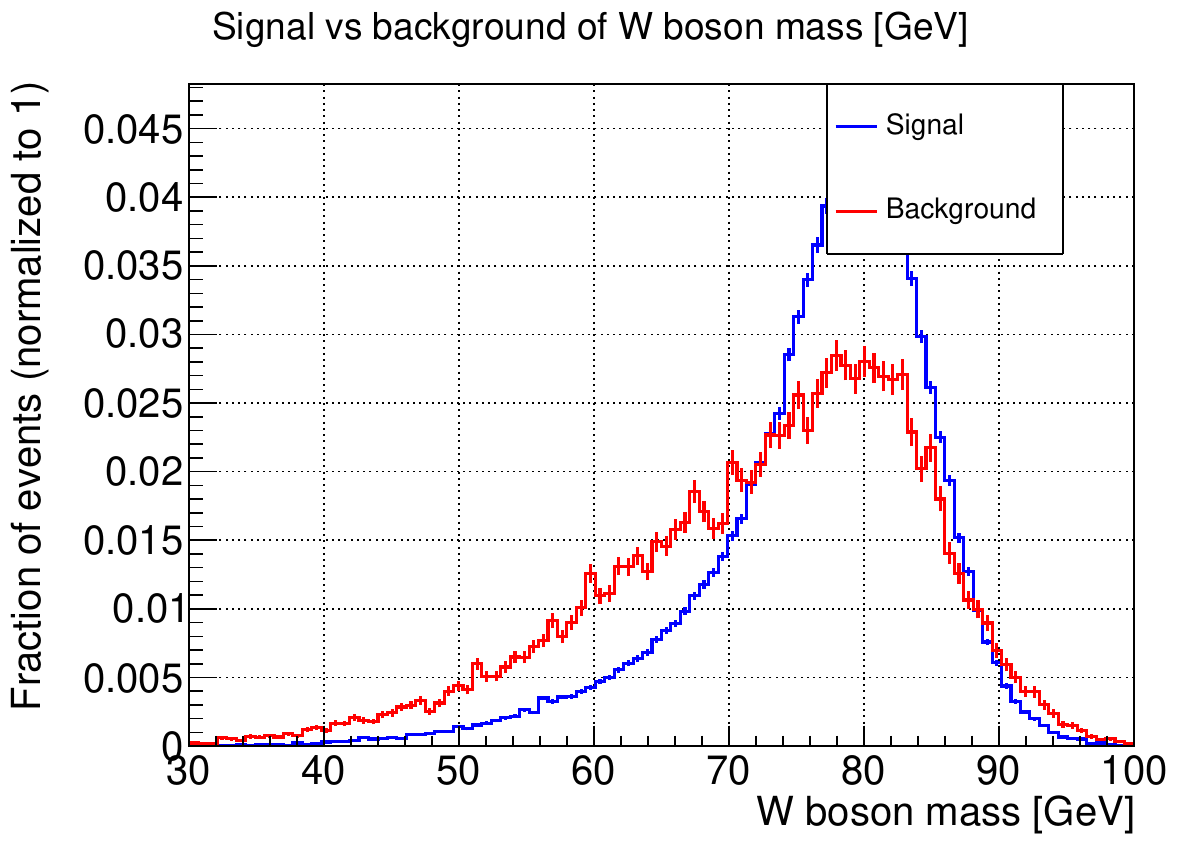}
    \end{subfigure}
    \caption{Distirbutions of the BDT input parameters not shown in \figref{fig:sig_bkg_bdt}. The figure format is the same as in \figref{fig:sig_bkg_bdt}.}
    \label{fig:sig_bkg_bdt_all}
\end{figure}

\subsection{Separate results for electron and muon channels}
\begin{figure}
    \centering
    \begin{subfigure}[t]{\linewidth}
        \centering
        \includegraphics[width=.9\linewidth]{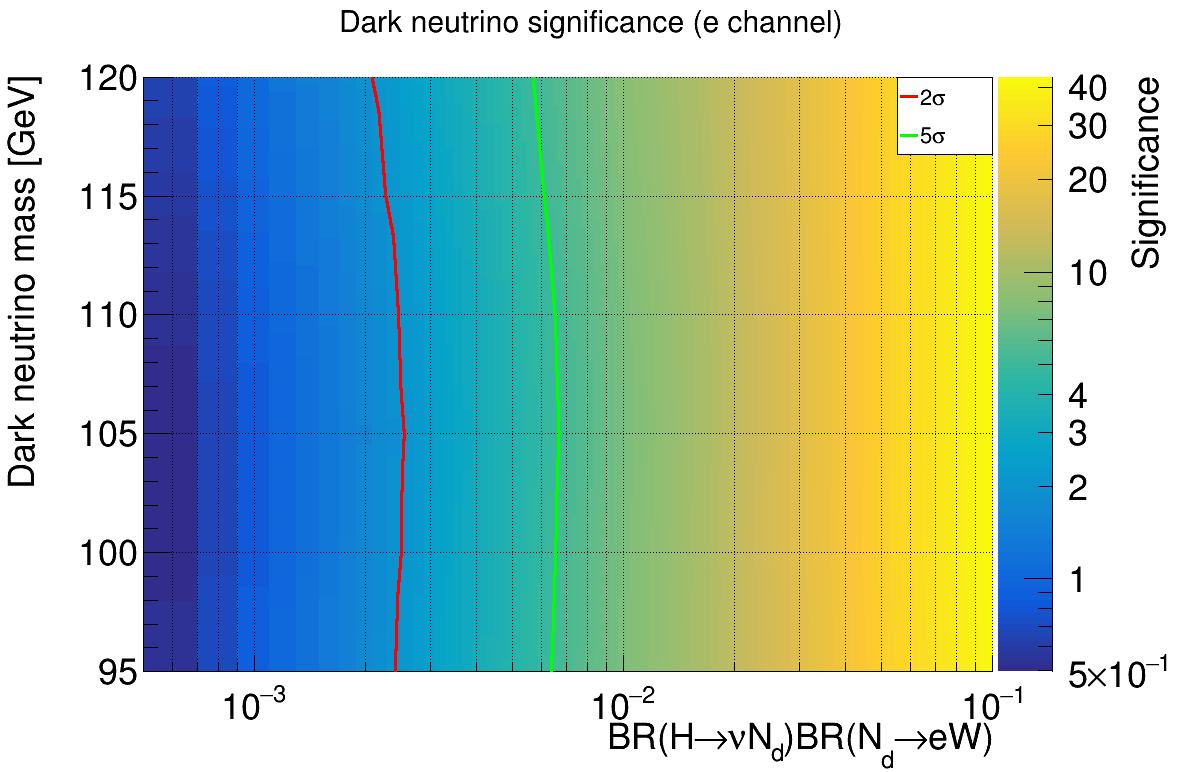}
    \end{subfigure}
    \hfill
    \begin{subfigure}[t]{\linewidth}
        \centering
        \includegraphics[width=.9\linewidth]{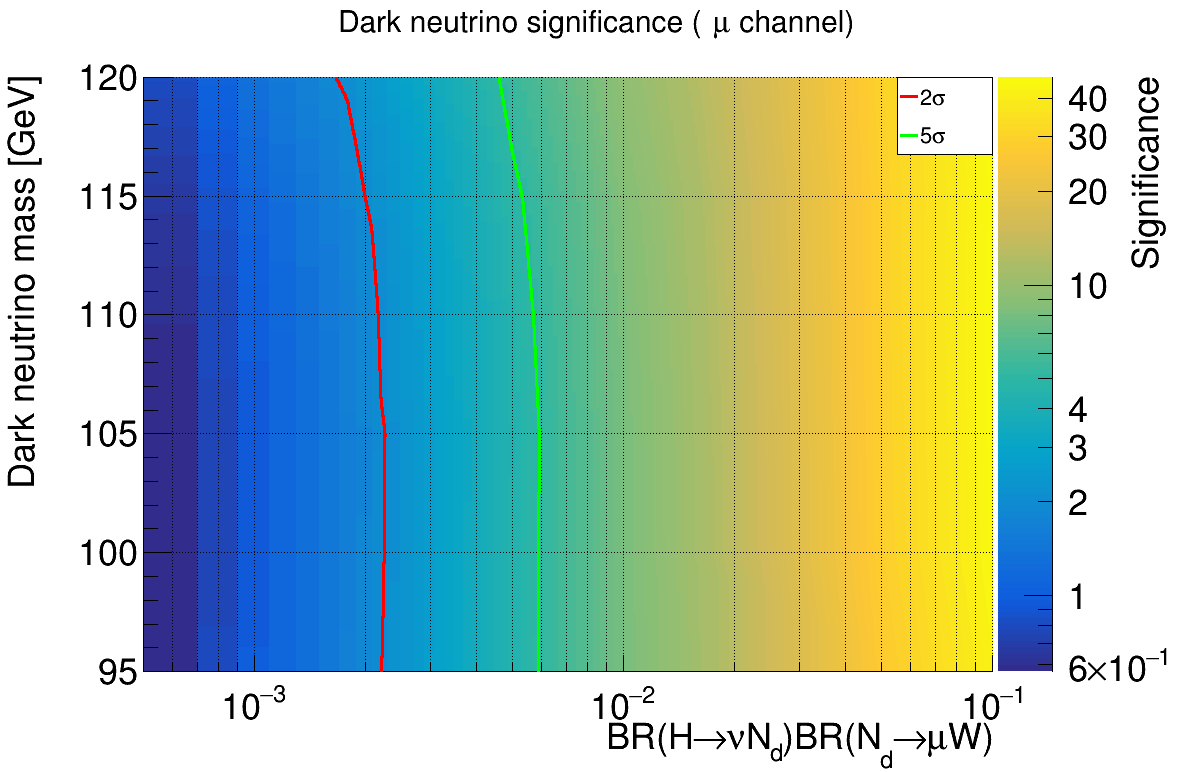}
    \end{subfigure}
    \caption{Exclusion plot as a function of branching ratio $BR(H\to \nu N_d)BR(N_d\to lW)$ ($x$ axis) and HNL mass ($y$ axis) for the two lepton channels. The top plot shows the electron channel ($N_d\to e^\pm W^\mp$) and the bottom plot shows the muon channel ($N_d\to \mu^\pm W^\mp$). The figure format is the same as in \figref{fig:exclusion}.}
    \label{fig:exclusion_emu}
\end{figure}

The significance for discovery is shown for both the electron and muon decay channel separately, i.e., $N_d \to e^\pm W^\mp$ and $N_d \to \mu^\pm W^\mp$, in \figref{fig:exclusion_emu}. The electron/muon separation is done by using the reconstructed particles and identifying them as electrons or muons.

\end{document}